# Anatomical basis of sex differences in the electrocardiogram identified by three-dimensional torso-heart imaging reconstruction pipeline

Hannah J. Smith[1], Blanca Rodriguez[1], Yuling Sang[2], Marcel Beetz[2], Robin P. Choudhury[3], Vicente Grau[2], and Abhirup Banerjee*[2,3]

[1] Department of Computer Science, University of Oxford, Oxford OX1 3QD, United Kingdom

[2] Institute of Biomedical Engineering, Department of Engineering Science, University of Oxford, Oxford OX3 7DQ, United Kingdom

[3] Division of Cardiovascular Medicine, Radcliffe Department of Medicine, University of Oxford, Oxford OX3 9DU, United Kingdom

[*] **Address for correspondence.**
Abhirup Banerjee, Institute of Biomedical Engineering, Department of Engineering Science, University of Oxford, Oxford OX3 7DQ, United Kingdom, abhirup.banerjee@eng.ox.ac.uk

## Abstract

The electrocardiogram (ECG) is used for diagnosis and risk stratification in myocardial infarction (MI). Women have a higher incidence of missed MI diagnosis and complications following infarction, and to address this we aim to provide quantitative information on sex-differences in ECG and torso-ventricular anatomical features and their interdependence. A novel computational automated pipeline is presented enabling the three-dimensional reconstruction of torso-ventricular anatomies for 425 post-MI subjects and 1051 healthy controls from UK Biobank clinical images. Regression models were created relating torso-ventricular and ECG parameters. We found that female hearts were positioned more posteriorly and superiorly than male, and in MI hearts were oriented more horizontally, especially for women. Post-MI women exhibited less QRS prolongation, requiring 27% more prolongation than men to exceed 120ms. Only half of the sex difference in QRS duration was associated with smaller female cavities. Lower STj amplitude in women was striking, associated with smaller ventricles, but also more superior and posterior cardiac position. Post-MI, T wave amplitude and R axis deviations were more strongly associated with posterior and horizontal cardiac positioning in women than in men. Our study highlights the need to quantify sex differences in anatomical features, their implications in ECG interpretation, and the application of clinical ECG thresholds in post-MI.

## Introduction

The electrocardiogram (ECG) is a vital tool for routine clinical assessment of cardiac electrical abnormalities. Patient differences in anatomy and electrophysiology both determine the ECG. Computational and experimental studies have demonstrated that torso-ventricular orientation can substantially affect the ECG, potentially confounding disease diagnosis, and anatomy rather than electrophysiology is the strongest contributor to ECG variability in healthy subjects [1-4]. Sex and body mass index (BMI) are strongly associated with differences in torso and heart anatomy, and thus it is crucial to take them into consideration in ECG interpretation [5-7]. ECG abnormalities are caused by alterations in the electrophysiological substrate, but importantly they are modulated by anatomical factors such as heart size, location and orientation with respect to the ECG leads [2, 3].

Ischemic heart disease is the leading single cause of mortality worldwide, killing 0.1% of the global population per year [8]. Therefore, effective patient diagnosis of myocardial infarction (MI) and risk stratification for treatment selection are crucial for survival and quality of life [9]. There is a higher incidence of missed diagnosis of acute MI in women, and hospital care can be less aggressive [10-12]. This is a potential contributor to the higher female mortality following MI [13], alongside factors such as differing demographic risk characteristics and comorbidities [14, 15].

Current guidelines for assessing arrhythmic risk post-MI use primarily mechanical markers and invasive testing [16]. They fail to identify the majority of sudden cardiac death cases, prompting calls for research into alternatives [17]. Several ECG-based biomarkers have been proposed building on mechanistic knowledge of the underlying electrophysiological substrate post-MI. QRS duration (QRSd) and its dispersion have been a key component of proposed electrocardiographic risk stratification following MI. However, findings of its effectiveness are mixed, with some trials suggesting it is a predictor of arrhythmic risk following MI [18, 19], while others showing it is more likely a predictor of general mortality than arrhythmic inducibility [20, 21]. T wave amplitude (TWA) and ST segment changes are also important, reflecting repolarisation heterogeneity [22]. Both of these biomarkers are included in current criteria for diagnosis and patient risk stratification in acute MI, with ST-elevation MI (STEMI) mandating immediate reperfusion therapy [23]. More recently, markers of repolarisation heterogeneity have also shown promise in risk stratification following MI [24, 25].

Women's ECGs typically exhibit shorter QRSd and lower ST junction (STj) and TWA, but longer QT intervals, than male [26]. Both electrophysiological and structural factors contribute to these differences. Testosterone increases repolarising potassium currents and decreases L-type calcium currents, both acting to speed up repolarisation [5]. Additionally, left ventricular (LV) mass is smaller in women than men [27], explaining in part the shorter QRSd and lower STj and TWA [28, 29]. Smaller torso volume can also increase amplitudes of the QRS and T waves in women [1, 2]. However, the impact on STj amplitude and whether anatomy is the only factor explaining these differences is unknown. In women, electrophysiological remodelling post-infarction may be less substantial than in men, partially from the protective effects of estrogen [30]. However, sex differences are often overlooked in the interpretation of ECG biomarkers, and thus investigations are urgently needed to fill these gaps for improvements in patient diagnosis, risk stratification, and treatment selection.

This study aims to quantify sex-differences in heart-torso anatomy and ECG biomarkers, as well as their relative associations, in healthy and post-MI subjects. To enable a population-wide study of heart-torso anatomy, we first develop and validate an automated image processing pipeline to reconstruct three-dimensional torso anatomical models using clinically standard cardiac magnetic resonance (CMR) images. We then apply the methodology to a cohort of 1476 UK Biobank (UKB) participants to statistically relate extracted anatomical parameters to relevant ECG biomarkers, including the effects of sex and MI status differences. We hypothesise that sex-differences in ECG biomarkers are only partially explained by smaller female heart and torso anatomies, and that sex-differences in electrophysiology and torso-heart position and orientation are particularly important, especially post-MI. These insights will lay the groundwork for a process to correct for the impact of torso-ventricular anatomy on the ECG, enabling personalised, automated ECG interpretation.

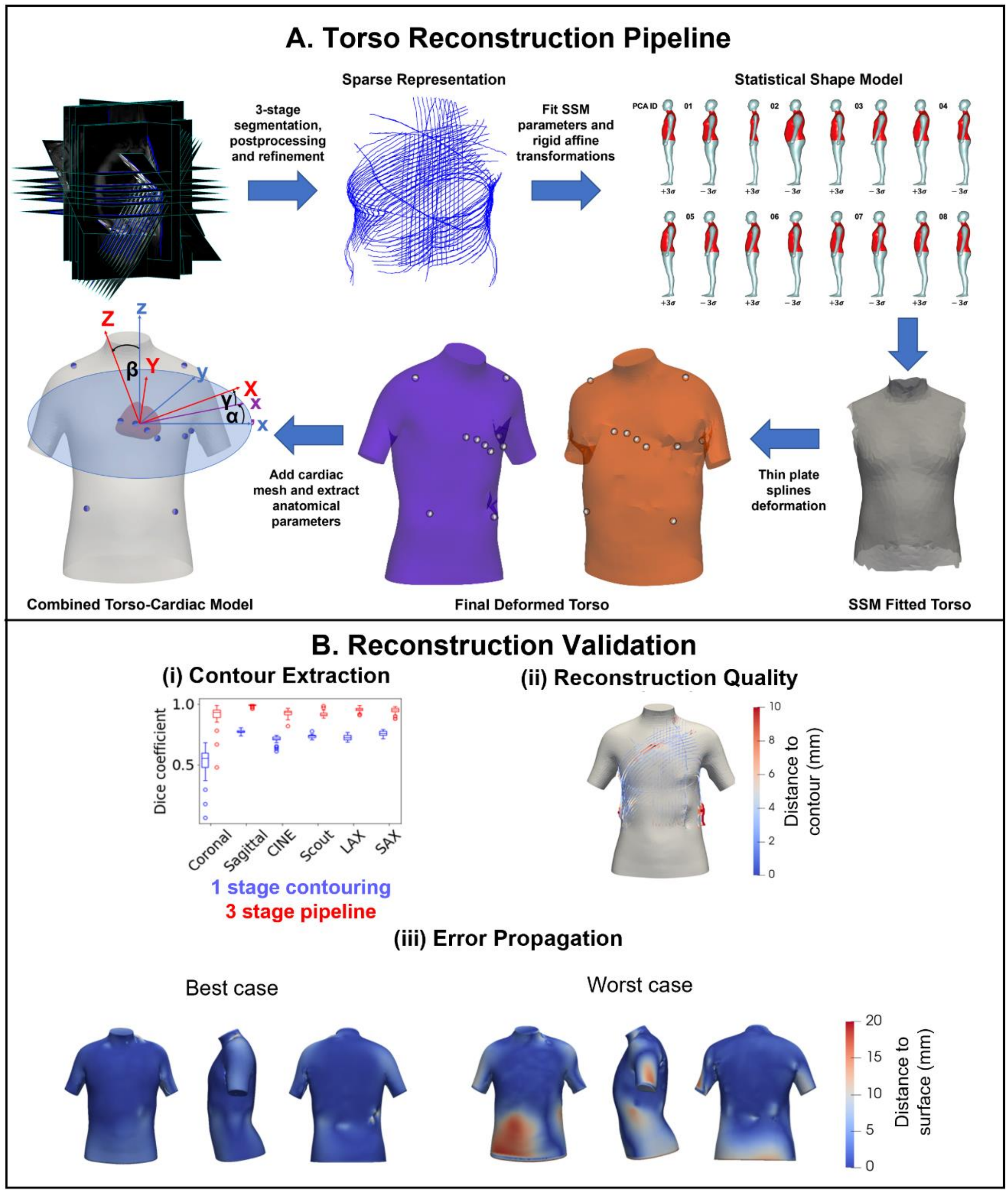

**Figure 1. Structure and validation of novel, automated torso reconstruction pipeline. A**: proposed end-to-end automated 3D torso reconstruction pipeline from 2D standard clinical cardiac magnetic resonance scans. The torso contours are first extracted from the images. A statistical shape model (SSM) is fitted to the contours, which is then optimally deformed. α: spin of the cardiac short axis plane around the torso vertical axis, β: verticality of the cardiac long axis, γ: tilt of the cardiac short axis plane. **B**: **(i)** comparison of Dice coefficient between single stage contouring (blue) and 3 stage segmentation, postprocessing and refinement (red). **(ii)** surface-to-contour distance for an example case between the automatically reconstructed surface and the automatically generated torso contours. **(iii)** surface-to-surface distance between the torso mesh created using the automated pipeline and the manually annotated contours for the subjects with the smallest (left) and largest (right) electrode error respectively.

# Methods

## Overview

A novel pipeline was first developed and validated to reconstruct torso anatomy from CMR imaging, which enabled the extraction of anatomical biomarkers describing the geometric relationship between the heart and torso. These anatomical biomarkers were then statistically related to relevant clinical ECG biomarkers in both healthy and post-MI populations in order to assess the effect of anatomical sex differences on the ECG.

## Computational pipeline

Figure 1A illustrates the fully automated pipeline used to generate personalised torso anatomical reconstructions from clinically standard CMR imaging. This integrates and reformulates state-of-the-art image processing techniques in order to give the first fully automated, and extensively validated, torso-ventricular reconstruction pipeline of its kind.

The pipeline combines machine learning-based segmentation and contour extraction, statistical shape modelling and automated mesh deformation in the following steps:

*Three-stage torso contour extraction from CMR imaging using machine learning to form sparse torso representation:* Approximately 60 localiser and scout images per subject were first segmented using a convolutional neural network, then their outline was extracted with automated post-processing, and finally refined using a second network. U-net [31] was used for both network architectures; detailed description of the architecture of these network, and the validation of the torso extraction (but not reconstruction) have been described in Smith et al. 2022 [32].

*Reconstruction:* To generate the initial 3D torso mesh, a statistical shape model (SSM) was applied over the automatically extracted sparse torso contours in 3D space, which was then optimally deformed by minimising the distance to the extracted contours to produce the final 3D torso mesh.

*Fitting the SSM*: In order to generate the torso meshes from sparse torso contours, a high-resolution SSM of human body shapes was employed for the initial 3D torso reconstruction. The SSM was generated from 3D optical body surface scanning of 4308 subjects in the CAESAR study, the largest commercially available scan database to date [33]. Detailed information on the SSM generation is provided in Pishchulin et al. 2017 [34]. As the initial condition, the mean SSM was translated into the subject's coordinate system by positioning the centre of heart from the mean SSM onto the centre of the subject’s heart. The latter centre was approximated by the closest point to the intersection lines between the cine short axis (SAX) and long axis (LAX) slices. Over the automatically extracted sparse torso contours in 3D space, the SSM was fitted by the optimal estimation of principal components, followed by rigid transformation, such that:

$$\min_{q,R} \sum_i d(C_i, R(\bar{X} + \Phi q))^2$$

where $C_i$ is the $i^{\text{th}}$ point on sparse torso representation, $\bar{X}$ the mean shape model, $\Phi = (\phi_1, \phi_2, \cdots)$ the SSM modes of variations, $q$ the estimated set of parameters along the principal modes, and $R$ the rigid transformation. The fitting of the reconstructions was restricted to the first 40 principal components of the SSM.

*Deforming for patient-specific torso reconstruction*: The initial generated torso mesh is inherently limited by the variability in the SSM. Hence, in order to accurately capture subject-specific variations, a final deformation step was performed via approximate thin plate splines (TPS) [35] to produce the final 3D torso mesh, minimising the distance to the extracted contours. In case the automated torso contour extraction generated sparse outlier points, any torso points over the 95th percentile were removed.

The initial deformation field was computed by considering the closest point on the SSM-generated mesh from the points on the sparse representation. A series of small, smooth deformations using approximate TPS was then applied to push the 3D mesh closer to the torso points, while preserving the smoothness and local topological properties [36]. Let F be the deformation field such that:

$$\arg\min_{F} \lambda \sum_{i} (||F(P_i) - Q_i'||)^2 + J_2^3(F) \text{ and } Q_i' = P_i + \beta \frac{(Q_i - P_i)}{||Q_i - P_i||}$$

where $\{P_i\}$ is the set of closest points on the surface $M_t$ to the torso points $\{Q_i\}$ on the sparse model, and $J_2^3$ is the TPS functional using derivatives of order 2 and image dimensions 3. The deformation $M_{t+1} = F(M_t)$ was iteratively applied in a diffeomorphic manner, resulting in a composition of several smooth approximations morphing initial mesh $M$ towards the sparse representation. Laplacian smoothing, decimation, and affine transformation were applied in the end to ensure the local geometric and topological characteristics of the reconstructed mesh. Finally, the torso surface was remeshed with a restricted Frontal-Delaunay algorithm using the mesh generator JIGSAW [37] with a specified element size of 1cm.

*Integration with a sparse cardiac mesh and subsequent extraction of torso-ventricular anatomical parameters:* Anatomical data for the size, position, and orientation of the heart were obtained in order to describe the geometrical relationship between the heart and torso. Cardiac surfaces were segmented using the method in Banerjee et al. 2021 [38] to give sparse cardiac anatomies, and thus calculate the cardiac size. Torso-ventricular anatomical parameters were extracted – full details for the calculation of each parameter are given in Supplementary Appendix 1.2. The torso volume and two metrics of cardiac size were estimated. The locations of the ECG electrodes were identified on the mean SSM torso, and their positions were transformed with the mesh such that the resulting torso had electrodes located in equivalent locations. As illustrated in Supplementary Figure 2, the position of the heart centre relative to the electrodes was estimated in the x (lateral), y (posterior) and z (superior) directions. Relative (dimensionless) parameters describing the percentage of the cardiac centre position between the most lateral and the most medial electrode were used in order to avoid the problem of collinearity with using multiple absolute distance measures, and similarly for the posterior and superior position. The orientation of the heart was quantified by the Euler angles of the cardiac axes with respect to the torso axes, shown in Supplementary Figure 3. This includes the spin of the cardiac short axis plane around the torso vertical axis, the verticality of the cardiac long axis, and the tilt of the cardiac short axis plane.

## Validation of the torso reconstruction pipeline

The CMR images of 30 subjects, chosen to be representative of the underlying dataset (as detailed in Supplementary Appendix 1.1), were manually annotated to form an independent test set for validation. Final results of the three-stage torso contour extraction process were compared with simply using a single u-net architecture to process raw images directly to refined contours in order to assess the accuracy of the torso-air boundary placement. Separate reconstructions of the 30 test subjects were made using the fully automated pipeline and using manually annotated contours. For each subject, the surface-to-surface distance between the two reconstructions was calculated to evaluate error propagation from contouring to reconstruction, as detailed in Supplementary

Appendix 1.3. The mean surface-to-contour distance was calculated for each test subject to evaluate the quality of the reconstruction.

## Dataset

The UKB dataset [39] was chosen for its high quality CMR imaging and 12-lead ECGs in a large population. 1646 control subjects were randomly selected from the UKB cohort alongside the 479 subjects with a history of MI, as defined in Supplementary Appendix 1.1. The healthy cohort was made by excluding control subjects with disease diagnoses, and subjects in either population were excluded if their ECG or imaging data was not complete, as described in Supplementary Appendix 1.1. The final population for reconstruction and statistical analysis comprised 1051 healthy and 425 post-MI subjects (total 1476), as shown in Table 1 and Supplementary Table 1. 92.7% of subjects identified as White British. The post-MI population was 6.1±0.4 years older and had 2.4±0.2 kg/m$^2$ higher BMI, with no statistically significant differences in ethnicity. This data was obtained through UK Biobank Resource under Application Number '40161' and the data was processed according to their guidelines. All participants gave written informed consent before enrolment in the UKB cohort.

**Table 1. Demographic characteristics of the 1476 subjects from the UK Biobank cohort.**

| Characteristics | Healthy (N = 1051) | MI (N = 425) | p-value |
|---|---|---|---|
| **Age (years)** | 61.1 ± 7.5 | 67.3 ± 6.2 | < 0.0001 |
| **Sex: Female, n (%)** | 581 (54.3) | 84 (19.8) | < 0.0001 |
| **BMI (kg/m$^2$)** | 25.9 ± 4.1 | 28.1 ± 4.4 | < 0.0001 |
| **Systolic blood pressure (mmHg)** | 133 ± 17 | 142 ± 19 | < 0.0001 |
| **Diastolic blood pressure (mmHg)** | 80.3 ± 10.0 | 81.4 ± 10.7 | 0.08 |
| **Ethnicity, n (%)** | | | 0.5 |
| White | 1021 (97.1) | 413 (97.2) | 1 |
| Mixed | 3 (0.3) | 1 (0.2) | 0.9 |
| Asian or Asian British | 8 (0.8) | 6 (1.4) | 0.2 |
| Black or Black British | 7 (0.7) | 1 (0.2) | 0.3 |
| Chinese | 5 (0.5) | 0 (0.0) | 0.2 |
| Other ethnic group | 5 (0.5) | 2 (0.5) | 1 |
| Prefer not to answer | 2 (0.2) | 2 (0.5) | 0.3 |

Continuous variables are shown as mean ± standard deviation and categorical variables as number of subjects (percentage in bracket). The p-value refers to the null hypothesis that the mean, or proportion in a category, is equal in the healthy and MI subpopulations. Distributions of continuous variables were tested for normality and equal variance, and the appropriate statistical test was selected, as outlined in Supplementary Appendix 1.4. $\chi^2$-tests were performed to assess the proportions of subjects for categorical variables. MI: myocardial infarction, BMI: body mass index.

## Statistical analysis

ECG parameters were obtained from the UKB database, which were automatically extracted with the proprietary software Cardiosoft [40]. QT and QTc interval were only available for lead I. The distribution of each anatomical and ECG parameter was compared between populations as in Supplementary Appendix 1.4. STj amplitude refers to the amplitude of the ST segment measured at the junction point (the end of the QRS complex). Each ECG parameter was separately linearly regressed against the set of anatomical parameters (chosen to minimise collinearity) for each subpopulation. This includes the cardiac and torso size, the relative position of the cardiac centre with respect to the electrodes in the lateral, posterior and superior directions and the three cardiac orientation parameters (spin, verticality, and tilt), all as previously defined. Additional linear models that also included the categorical sex parameter were obtained for mixed sex populations, and multiplied by the difference in the parameter between men and women to estimate their contribution to the ECG biomarker sex difference. The effect of electrophysiology was estimated as the coefficient of the sex variable after all anatomical parameters had been controlled for. The contributors to the difference between healthy and post-MI populations were similarly calculated, separately for men and women. A linear model was trained to estimate, and therefore correct, the total anatomical contribution to ECG parameters using only demographic features (sex, height, BMI, and age), as described in Supplementary Appendix 1.4.

# Results

## Reconstructed torso surfaces matched input contours to within 2.19mm:

Figure 1B illustrates the validation of the torso reconstruction pipeline, showcasing its capability to extract torso contours and reconstruct them into 3D meshes. The three-stage segmentation, automated post-processing, and refinement procedure improved the Dice coefficient of torso contour extraction, compared to a single-stage contouring network, across all image views (Figure 1B(i)). The median surface-to-contour distance, showing the quality of the reconstructed torso surface (Figure 1B(ii)), for the test subjects was 2.19mm (inter-quartile range - IQR: 1.98-2.37mm). The median surface-to-surface distance between torso meshes reconstructed using the manually annotated contours and the machine learning pipeline for the 30 test subjects was 0.82mm (IQR: 0.69-1.10mm). Figure 1B(iii) shows the reconstructed surfaces for the test subjects with the smallest and largest median reconstruction errors. Both reconstructed surfaces were smooth and anatomically realistic, and, across most of the torsos, errors were within the millimetre range.

Torso volume was larger in post-MI than in healthy subjects in women and men, by 3.7±0.9 $dm^3$ (10.2±3.3%) and 3.5±0.5 $dm^3$ (8.2±1.9%) respectively (Supplementary Figure 4). Additionally, for women, in post-MI the heart was positioned more posteriorly by 3.1±0.9mm, than in healthy subjects. The heart was also oriented with its long axis more horizontally by 6.0±1.2° and 4.3±0.5° in post-MI versus healthy women and men, respectively. These changes were related to the increased BMI of post-MI subjects (Supplementary Figure 4K). Age was also considered (Supplementary Figure 4J), but its effects were considerably less substantial.

## QRS duration was shorter in women than in men, particularly post-MI, due to smaller hearts and reduced impact of electrophysiological abnormalities:

For both healthy and post-MI subjects, QRSd was shorter in women versus men in all leads (Figures 2A and 2B), with contributions from each anatomical factor displaying sex differences (Figures 2C and 2D). Sex differences in cavity volume explained approximately half of the sex difference in QRSd in both healthy and post-MI subjects (a mean value across all leads of 3.4±1.3ms out of the 6.0±1.5ms mean total in healthy, and 4.5±1.4ms of a total sex difference of 8.3±2.5ms in post-MI). In healthy subjects, the electrophysiological contribution to QRSd sex differences was similar in all precordial leads (1.0±1.0ms to 2.3±1.2ms). In post-MI subjects, it was considerably more variable (0.2±2.8ms in V1 to 6.3±3.0ms in V6). The total estimated impact of anatomy on the QRSd was over 5ms in lead V6 for 12.4% of healthy subjects. After correction by a linear model that was trained only on demographic features (sex, height, BMI, and age), this proportion reduced to 2.9%.

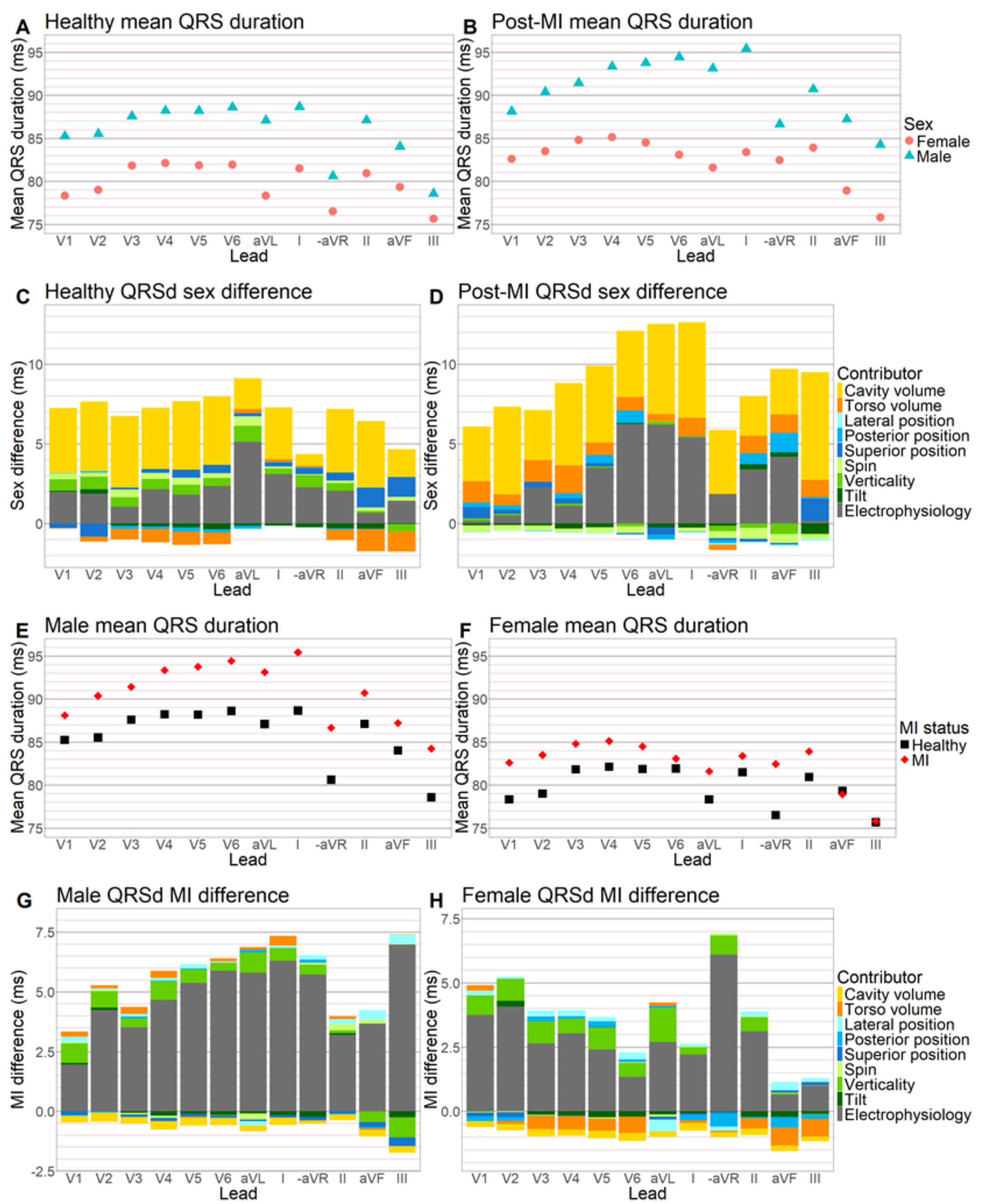

**Figure 2. Sex differences in QRS duration in healthy and post-MI subjects. A, B:** mean QRS duration (QRSd) for each ECG lead in healthy and post-MI subjects respectively with women (red circles) and men (cyan triangles). **C,D:** contribution of anatomical parameters and electrophysiology to sex differences in QRSd for healthy and post-MI subjects (calculated by multiplying the regression coefficient for each factor by its mean difference between male and female populations). **E**, **F:** mean QRSd for each ECG lead in men and women respectively for healthy (black squares) and post-MI (red diamonds) subjects. **G, H:** contribution of anatomical parameters and electrophysiology to differences in QRSd between healthy and post-MI subjects for male and female subjects (calculated by multiplying the regression coefficient for each factor by its mean difference between healthy and post-MI populations).

Post-MI women exhibited less QRS prolongation than men (Figures 2E and 2F), with mean percentage increase in QRSd from healthy to post-MI, across all leads, of 3.4±2.3% and 5.8±1.5% respectively. This is explained by a reduced contribution of electrophysiological abnormalities post-MI in women, as shown in Figures 2G and 2H (smaller grey bars).

The QT interval was 6.3±2.0ms and 1.1±3.9ms longer in healthy and post-MI women than men respectively in lead I (the only lead with available QT interval data). Figure 3A shows the electrophysiological and anatomical contributions to these differences. While electrophysiological differences prolongated the female QT interval in comparison with male (shown as negative gray contributions), this was partially compensated by the smaller female ventricles (resulting in shorter female QT interval). Figure 3B shows the contributions to the 15.2±1.6ms and 11.0±3.3ms longer QTc interval in healthy and post-MI women respectively than in male. Importantly this shows that the anatomical contribution was substantially reduced – for example from 18.9±5.9ms to 2.5±2.1ms in the healthy population, primarily by near elimination of the contribution from the cardiac size. Additionally including heart rate in the QRSd correction model reduced the proportion of cases where corrected anatomical contribution exceeded 5ms from 2.9% to 2.3%.

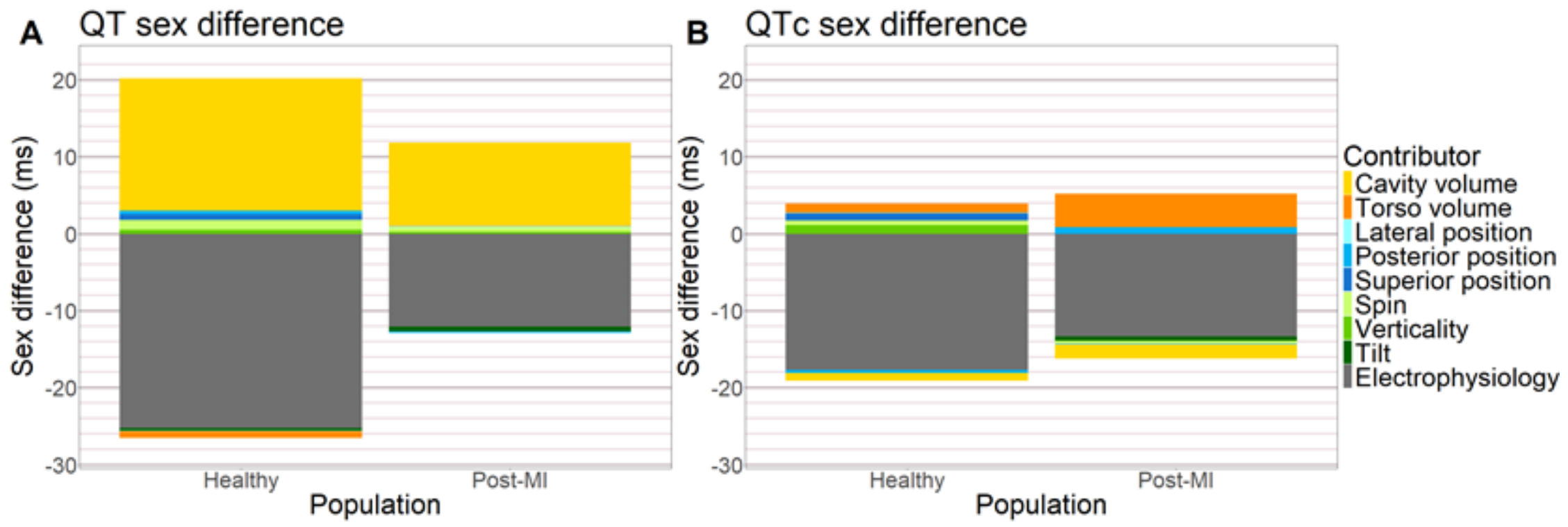

**Figure 3. Sex differences in raw versus corrected QT interval.** Contribution of anatomical parameters and electrophysiology to sex differences in raw QT (**A**) and corrected QT (**B**, QTc) intervals for healthy and post-MI subjects in lead I (calculated by multiplying the regression coefficient for each factor by its mean difference between male and female populations).

## ST junction amplitude was lower in women than men in both healthy and post-MI subjects, explained by smaller hearts in more superior and posterior position:

STj amplitude was lower in women versus men, particularly in the septal and anterior leads, in healthy and post-MI subjects (Figures 4A and 4B). As for QRSd, this was partly explained by women's smaller LV mass (by 29.3±5.4% and 27.1±5.7% in healthy and post-MI subjects respectively), but also by a more superior position in women than men for healthy subjects (by 9.6±0.9mm) and more posterior position for post-MI subjects (by 4.3±1.0mm) (Figures 4C and 4D). Both smaller LV mass and electrophysiology contributed to lower female STj amplitude, whereas a smaller female torso volume was generally associated with a higher STj amplitude (yellow and grey versus orange bars in Figures 4C and 4D). This pattern was consistent for differing measurement times of the ST level (Supplementary Figure 8). Baseline STj amplitude in lead V4 was 0.028±0.003mV lower in healthy women than healthy men. Healthy women with BMI>25kgm$^{-2}$ had a STj amplitude 0.047±0.004mV lower than healthy men with BMI<25kgm$^{-2}$, associated with their smaller ventricles, but larger torsos.

## Lower T wave amplitude following MI was more strongly associated with a more posterior cardiac position and more horizontal orientation in women than men:

Precordial TWA was lower in women than men, by 41.7±20.3% and 31.2±20.0% in healthy and post-MI populations respectively (Supplementary Figure 9). MI resulted in reduced TWA, in all leads in men, whereas only in V4-6 and limb leads in women (Figures 4E and 4F). For men, this was largely associated with electrophysiological factors substantially reducing TWA across leads (Figure 4G). However, for women, the reduced precordial TWA was largely associated with more posterior and horizontal ventricles from their larger BMI (Figure 4H), and electrophysiological factors actually counteracted the TWA post-MI reduction for septal leads (shown as positive contributions).

## R axis deviations following MI were associated with more horizontal cardiac orientation in women, but MI-related electrophysiological abnormalities in men:

Figures 5A and 5B break down the anatomical contributions to R and T axis shifts in post-MI compared with healthy subjects. For men, the electrophysiological contribution dominated their 12.0±2.8° left deviated R axis following MI (Figure 5A). However, for women their 5.3±4.0° left deviation in the R axis was mainly associated with a more horizontal cardiac long axis, with electrophysiological factors playing very little role (Figure 5B). This was partly explained by female R axes being more sensitive to changes in verticality – the regression coefficient between R axis and verticality was higher for women than men (Figure 5C). The female T axis was more right deviated following MI than male, largely associated with electrophysiology (Figures 5A and 5B), with less contribution from orientation (Figure 5D).

Torso volume significantly affected R axis for both sexes in healthy, but not post-MI subjects (Figure 5E, regression coefficient of R axis against torso volume). For example, in healthy women, every $10dm^3$ increase in torso volume was associated with a decrease in R axis of 7.2±3.0°. This was related to an increased distance to the left leg electrode (inversely related to the R axis) without as substantial an increase in distance to the left arm electrode (positively related to the R axis). For healthy women, every $10dm^3$ increase in torso volume was associated with an increased distance to the heart centre of 19.4±0.7mm and 5.9±0.6mm for the left leg and arm electrodes, respectively. However, in post-MI women, the gap between these differences was smaller, with corresponding increased distances of 16.6±1.8mm and 8.3±1.7mm. This association was less strong for the T wave (Figure 5F).

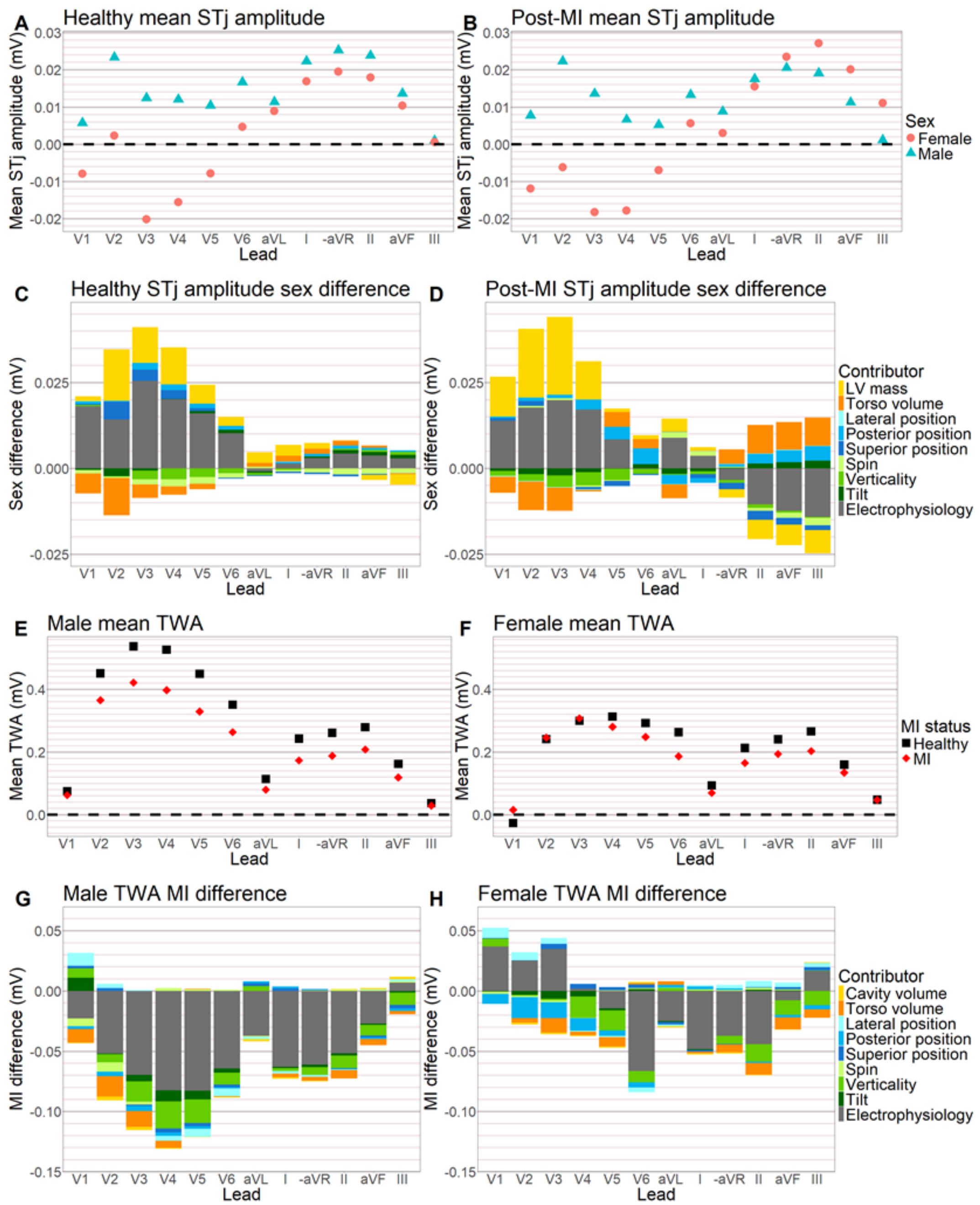

**Figure 4. Sex differences in ECG amplitudes in healthy and post-MI subjects. A, B:** mean STj amplitude (measured at QRS offset) for each ECG lead in healthy and post-MI subjects respectively for women (red circles) and men (cyan triangles). **C, D:** contribution of anatomical parameters and electrophysiology to sex differences in STj amplitude for healthy and post-MI subjects (calculated by multiplying the regression coefficient for each factor by its mean difference between male and female populations). **E**, **F:** mean T wave amplitude (TWA) for each ECG lead in men and women respectively for healthy (black squares) and post-MI (red diamonds) subjects. **G, H:** contribution of anatomical parameters and electrophysiology to differences in TWA between healthy and post-MI subjects for male and female subjects (calculated by multiplying the regression coefficient for each factor by its mean difference between healthy and post-MI populations).

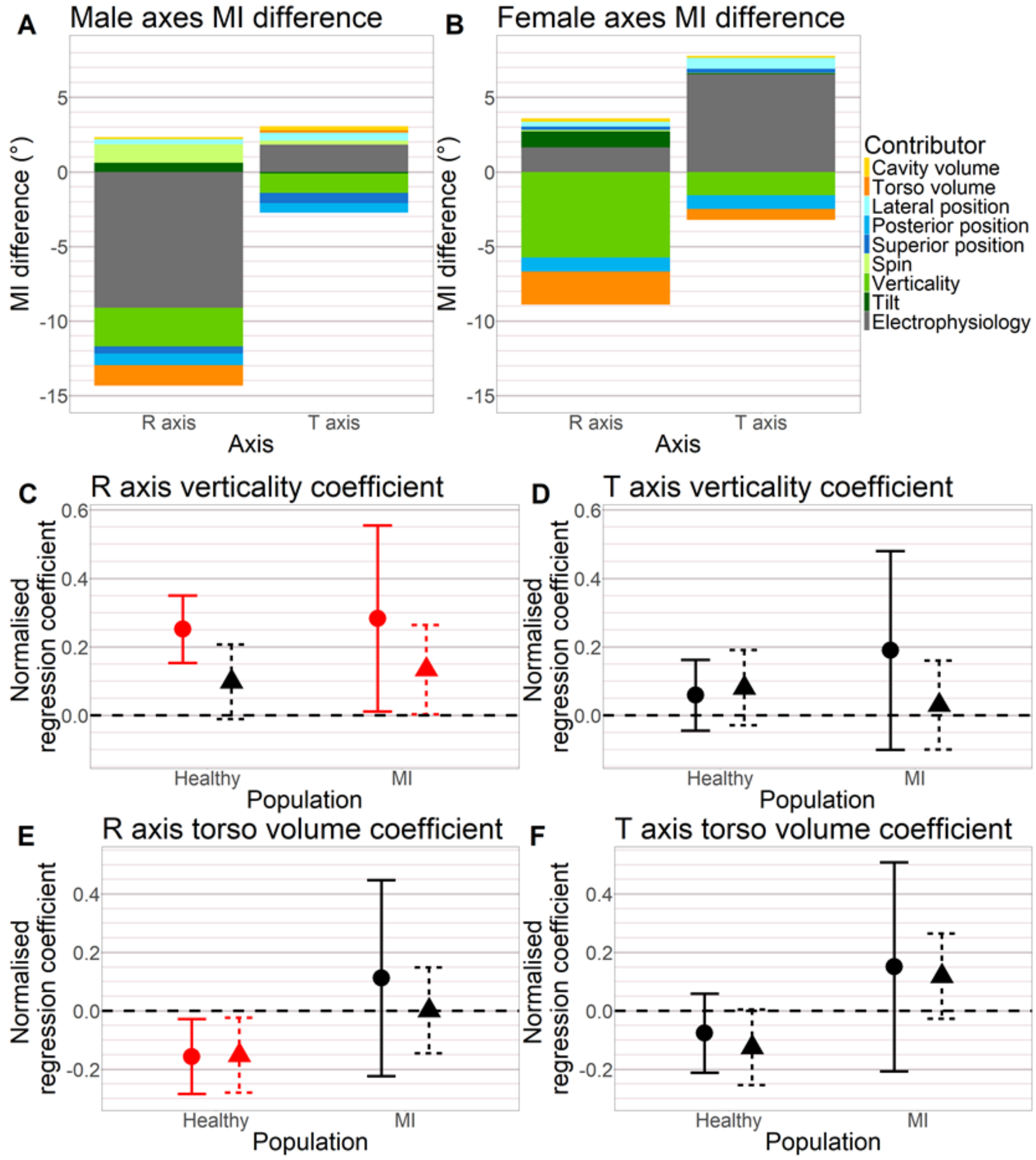

**Figure 5. Sex differences in ECG axis angles in healthy and post-MI subjects A**, **B:** contribution of anatomical parameters and electrophysiology to differences in R and T axis angles between healthy and post-MI subjects for male and female subjects (calculated by multiplying the regression coefficient for each factor by its mean difference between healthy and post-MI populations). **C**, **D:** normalised regression coefficients showing the association between the verticality of the cardiac long axis and the R and T axis angles respectively for the healthy and post-MI populations. Bars shown in red represent regression coefficients significantly different from 0, at a significance level of 0.05. **E, F:** normalised regression coefficients showing the association between torso volume and the R and T axis angles respectively for the healthy and post-MI populations.

## Discussion

This study provides quantification of sex differences in image-based anatomical factors and their implications for ECG biomarkers in healthy and post-MI subjects, enabled by a novel computational clinical image processing pipeline. With this work, we aim to contribute to addressing the higher incidence of missed diagnosis and increased female MI mortality following infarction [12, 13, 41]. Our analysis of a large cohort of healthy and post-MI subjects' images and ECG data provides quantitative evidence on the importance of considering demographic characteristics and their anatomical differences, particularly sex, for corrections of ECG biomarkers, key to patient risk stratification.

An important contribution is the novel computational pipeline for automated 3D torso reconstruction, which enables high-throughput investigations into the effects of anatomical variability on the study of a wide range of cardiac diseases. The adaptation of the pipeline to standard clinical cardiac imaging allows for the exploitation of large databases, such as the UKB, and sets the groundwork for clinical tools to personalise ECG interpretation considering patient anatomy.

Our analysis provides the following insights into sex-related anatomical differences of the impact of MI on key ECG biomarkers: 1) Half of the shorter QRSd in women was explained by smaller cavity volumes in both healthy and post-MI subjects. 2) Female STj amplitude was lower than male in all precordial leads for healthy and post-MI subjects, associated with a smaller LV mass and more superior and posterior position. 3) In women, anatomical factors had a stronger impact on TWA changes following MI than in men. 4) R axis was more left deviated in MI than healthy populations; for women this was associated with a more horizontal cardiac long axis, but for men with electrophysiology.

Both clinically and moderately prolonged QRSd have been associated with increased risk following MI [18, 19]. Importantly, our analysis reports that mean QRSd would have had to increase by 27% more for women than men to exceed 120ms, and by 68% more to exceed 100ms, both in lead aVL. Sex differences in QRSd are only partly explained by cardiac size, in line with Mincholé et al. 2019 [1], and with the biophysical understanding that larger distances require longer to traverse [42]. However, even after accounting for anatomical factors, there seems to be an electrophysiological component to the increased male QRSd, which supports previous works finding that men had more complex activation pathways, prolonging their QRSd [43]. Our finding showed that taking into account only simple demographic information (age, sex, BMI, and height) substantially reduced the impact of anatomy on QRSd. This suggests that correction methods without complex reconstruction may be sufficient to capture anatomy-ECG relationships in MI diagnosis and risk stratification. Nonetheless, detailed characterisation of cardiac shape utilising 3D reconstruction has been shown to provide additional prognostic value and improve risk stratification following infarction [44, 45]. QRS prolongation in post-MI subjects has been reported in previous studies, particularly cooccurring with ST changes in leads exploring the affected region [46]. It can be explained by slowed conduction velocity in the affected region, involvement of the conduction system such as the bundle branches and Purkinje, and in the chronic case, with dilation and hypertrophy of non-infarcted myocardium [20, 47, 48]. The fact that the pattern across leads of the electrophysiological prolongation of QRSd following MI differed between men and women potentially points to differences in the location or direction of those conduction disturbances. Our finding that sex differences in QTc had much less contribution from cardiac size than in QT suggests that part of the anatomical masking of the longer QT interval in women is because their smaller hearts naturally beat faster, shortening the QT interval. There is some evidence that QRS duration also shortens with faster heart rate, though to a lesser extent, and the heart rate may serve as an easily obtained proxy for cardiac size, so correction by heart rate may play a role in QRSd-based risk stratification [49].

The finding that STj amplitude was substantially lower in women than men in all septal and anterior leads (V1-V4) is also striking. Based on our findings, the dual effects of cardiac and torso volume need to be considered in the interpretation of ST-elevation through the lens of the anatomical context of the patient. As we considered healthy and post-MI subjects (rather than with acute ischemia), our data reflect the normal baseline of STj amplitude, from which any deviations in acute ischemia can be measured. The Fourth Universal Definition of Myocardial Infarction only has sex-specific thresholds in leads V2 and V3 for identifying ST-elevation, a key component in the diagnosis of STEMI [50]. For example, in lead V4, the clinical threshold for both female and male ST-elevation is 0.1mV [50]. Based on our findings, due to their lower baseline values, overweight healthy women would need a 63% larger increase in STj amplitude in this lead to be classified as ST-elevated than normal weight healthy men. Therefore, this work represents a significant step forward in the characterisation of anatomical influences of its baseline level, from which deviations can then be more accurately identified in the context of anatomical variation. It also highlights the important intersectional effects of demographic characteristics such as sex and BMI, suggesting that both of these factors should be accounted for in assessing ST-elevation to improve accuracy of MI diagnosis and classification. We found a substantial electrophysiological component to the sex difference in STj amplitude in both healthy and post-MI subjects. This potentially happens as the longer female action potential duration causes less overlap between activation and early repolarisation in different sections of tissue occurring at the same time [51]. Again, as this occurred in leads without sex specific MI diagnostic guidelines such as V1 and V4, such thresholds should be reconsidered.

Alterations to the polarity and amplitude of the T wave are used in diagnosis of acute MI [50] and TWA affects proposed risk stratification tools, particularly markers of repolarisation abnormalities [9, 52]. Our data reveal that, particularly in women, a substantial proportion of TWA reductions were associated with anatomical differences, such as the decreased verticality of the post-MI heart. This implies that without anatomical corrections, such as adjusting for BMI, the female TWA may be less representative of its underlying electrophysiology, and therefore less representative of any pathology. This increased effect of anatomical parameters in women was partly due to their larger difference in key anatomical factors such as posterior position and long axis orientation from healthy to post-MI subjects. This reflects their larger increase in BMI, associated with a more horizontal cardiac long axis, as described in previous works, suggesting this was related to an upwards shift in the diaphragm [6, 53]. Furthermore, the female TWA was more sensitive to changes in anatomical parameters. This could be explained by sex differences in the directions of the T wave vectors and nonlinearities in the structure-function relationships. Our finding that TWA was affected by cardiac position and orientation is in line with previous simulation studies [2] albeit in a small number of anatomies (N=5) and without electrophysiological variation. Electrophysiologically, TWA variations are caused by differences in the magnitude and timing of repolarisation gradients, and lower TWA in healthy women than men is in part related to their slower repolarisation [54]. TWA drops, and even inverts, in MI due to disruption of the usual direction of the transmural repolarisation gradient from adverse remodelling, alongside tissue death [55]. Our results found a different pattern of electrophysiologically-driven TWA changes following infarction between men and women. This could point to either differences in the nature of the repolarisation remodelling, or how this remodelling interacts with naturally present sex differences in cellular electrophysiology.

Our data revealed that the verticality of the cardiac long axis was particularly significant in determining R axis angle, especially for women. This represents a step forward in understanding the complex relationship between electrical and anatomical axes [6, 56, 57]. Moreover, the left-deviation of the R axis following infarction may look similar in both sexes; in men this was representative of an electrophysiological change, presumably alterations to the direction of the QRS vector caused by both hypertrophy and areas of electrical inactivity [47]. However, for women this change was almost entirely associated with anatomical differences. This suggests that it is particularly important to correct R axis angles for anatomical factors in women to avoid misattribution of deviations to electrophysiological pathology. The significant relationship between torso volume and R axis angle for healthy subjects indicates that BMI and height may be critical considerations when using ECG based biomarkers that are sensitive to axis angles. The fact that increased torso volume moved the left leg electrode further away from the heart centre than the left arm electrode suggests that this relationship is caused by a stretching of the anatomical plane in which these angles are viewed. Therefore, this represents another area in which a clinical tool that accounts for anatomical variation in interpreting ECG biomarkers would reduce differential accuracy in risk stratification between demographic groups.

The computational pipeline presented here is available to be applied to further studies. Validation was performed by demonstrating that the extracted contours are near the ground truth torso-air boundary on the clinical imaging, as well as to the final reconstructed surfaces (i.e. high Dice coefficient for contour extraction and low surface-to-contour distances). As detailed in Smith et al. 2022 [32], the three-step segmentation, automated post-processing, and refinement process was particularly effective at removing errors inside the torso at locations of sharp gradient between dark and light, for example due to air in the lungs. Whilst reconstructed 3D torso surfaces were largely smooth and realistic, there were small irregularities, generally on the shoulder and waist regions. These were not typically associated with incomplete CMR artifact removal and may instead result from images being taken on different breath-holds or the image sparsity. The low surface-to-surface distance between reconstructions made using the automatically and manually derived contours demonstrates that the small errors in contour extraction did not lead to significant errors in electrode placement on the torso meshes. Surfaces were generally more accurate on the upper left quadrant of the torso, which is where most electrodes are placed. The validity of the torso-ventricular reconstruction pipeline is further supported as known trends in torso and cardiac size, position, and orientation were qualitatively reproduced, and furthermore were quantitively characterised. This includes the more superior female cardiac position [58, 59] and the horizontal shift in the cardiac long axis for subjects with a larger BMI [6, 53]. As this pipeline uses clinically standard CMR imaging, it can be more easily translated to a clinical tool to correct for the effect of anatomical variation on the ECG in a variety of medical contexts.

Due to the demographics of the UKB [39], this study only included subjects aged 45-80 years old; however, the incidence of MI is very low in people younger than this range [13]. The ethnic background of the dataset also had limited diversity; however, this is roughly representative of the UKB population and thus the ethnic composition of the United Kingdom at the time of recruitment [60]. As databases become more diverse, the mediating effect of ethnic background on the cardiac structure-function relationship should be investigated. Whilst the healthy subjects had a roughly equal sex balance, the post-MI subjects had a male-female ratio of approximately 4:1. Whilst this is in line with sex differences in diagnosed MI events [61], the proportion of female MI events may be underrepresented due to more missed diagnoses, particularly for more mild cases [12]. This may have skewed the diagnosed female post-MI ECG phenotype to be more severely affected, and the demographics to be, for example, older. Taking anatomical sex differences into account when diagnosing acute MI may aid in addressing this imbalance in future MI investigations.

The statistical shape model used for initial torso reconstruction did not have separate male and female models, which may have led to an underestimation of the sex differences in the torsos of male and female anatomies. However, the deformation step of the reconstruction was designed to address this issue by modelling personalised torso variations. The exact electrode positions of the obtained ECGs were not available in the UKB dataset, so these were estimated by placing the electrodes on standard positions on the personalised torso. However, this estimation process allows for use of many similar clinically standard datasets, avoiding complex acquisition protocols. Both voluntary and involuntary patient movement impedes precise determination of the cardiac positioning and the reconstruction of a static torso surface, and is one factor explaining remaining reconstruction errors. The large number of subjects in this study likely reduced the impact of this noise on the statistical results.

As the anatomical factors cannot represent a perfectly complete and independent basis set, the electrophysiological contribution can only be estimated. However, as there is more anatomical variation than can be represented using the given factors, such as the smaller scale geometry of the heart, the role of the anatomy versus electrophysiology was likely under- rather than overestimated. Computational modelling and simulation techniques [62, 63] are well suited to further investigate the impact of electrophysiological sex differences on the ECG in future works. Whilst the study benefits from the electrophysiological and anatomical variation that using clinical data facilitates, it is impossible to fully separate the effects of these factors on ECG biomarkers, as characteristics such as sex affect both categories, making them interdependent. This means that this work complements computational studies, which have less interpersonal variability, but isolate the effect of positional and rotational changes. Scar characteristics were not available in the UKB, but the effect of scar geometry variation could be considered in future studies with the development of scar delineation tools. Future work should focus on the characterisation of ECG biomarkers in MI and other disease conditions. The automated torso reconstruction pipeline this work proposes enables the expansion of the number of anatomies such future computational studies could investigate.

# Conclusion

To conclude, this work presents a novel automated pipeline for personalised torso reconstruction and demonstrates its power in exploiting large clinical databases by relating torso-ventricular anatomy and ECG biomarkers for 1051 healthy and 425 post-MI subjects. Results show the considerable influence of anatomical factors in demographic differences of ECG biomarkers. Women and individuals with higher BMI may be disproportionately affected. This underlines the importance of, and lays the foundations for, personalised ECG interpretation that considers a subject's individual torso-ventricular anatomy, facilitating improvement in diagnosis and risk stratification tools.

## Funding

The work of HJS was supported by the Wellcome Trust grant 102161/Z/13/Z. For the purpose of open access, the authors have applied a CC BY public copyright licence to any Author Accepted Manuscript version arising from this submission. AB is supported by the Royal Society University Research Fellowship (Grant No. URF\R1\221314). The works of AB, RPC, and VG are supported by the British Heart Foundation (BHF) Project under Grant PG/20/21/35082. This work was funded by a Wellcome Trust Fellowship in Basic Biomedical Sciences to BR (214290/Z/18/Z) and the CompBioMed 2 Centre of Excellence in Computational Biomedicine (European Commission Horizon 2020 research and innovation programme, grant agreements No. 823712) to BR and VG.

## Acknowledgements

This research has been conducted using the UK Biobank Resource under Application Number '40161'. The authors would like to thank Dr Ernesto Zacur for providing valuable suggestions on the 3D anatomical mesh reconstruction.

## Additional information

The authors have no competing interests.

## Data availability

The UKB data is publicly available and can be accessed by application upon approval here: https://www.ukbiobank.ac.uk/enable-your-research/register. The reconstruction pipeline code is publicly available here: https://github.com/MultiMeDIA-Oxford/TORSO-MPP, with explanatory documentation for its use.

**Author contributions statement.** HJS, BR, VG, and AB conceptualised the study and designed the research. BR, VG, RPC, and AB supervised the research. HJS and AB acquired and curated the dataset. AB and HJS developed the reconstruction pipeline. HJS, AB, YS, and MB pre-processed and cleaned the data. HJS and AB conducted the study and analysed the data and the results. HJS, AB, BR, and VG prepared the original draft and reviewed and edited the manuscript. All authors discussed the results, provided comments regarding the manuscript, and agreed on the final draft.

# Supplemental Materials of "*Anatomical basis of sex differences in the electrocardiogram identified by three-dimensional torso-heart imaging reconstruction pipeline*"

Hannah J. Smith[1], Blanca Rodriguez[1], Yuling Sang[2], Marcel Beetz[2], Robin P. Choudhury[3], Vicente Grau[2], and Abhirup Banerjee[2,3]

[1] Computational Cardiovascular Science Group, Department of Computer Science, University of Oxford, Oxford OX1 3QD, United Kingdom

[2] Institute of Biomedical Engineering, Department of Engineering Science, University of Oxford, Oxford OX3 7DQ, United Kingdom

[3] Division of Cardiovascular Medicine, Radcliffe Department of Medicine, University of Oxford, Oxford OX3 9DU, United Kingdom

[*]**Address for correspondence.**

Abhirup Banerjee, Institute of Biomedical Engineering, Department of Engineering Science, University of Oxford, Oxford OX3 7DQ, United Kingdom, abhirup.banerjee@eng.ox.ac.uk

## Supplementary Appendix 1: Methods

**1.1 Dataset.** Myocardial infarction (MI) was defined as any history of MI including ST- and non-ST-elevation MI, as obtained from patients' self-report in the baseline questionnaire and nurse-led interview, and linked hospital admission data [64]. The exclusion criteria for the subjects used in the larger statistical analysis are shown in Supplementary Figure 1. From the 1646 controls, 3 cases were excluded for incomplete image sets, and 592 were excluded for having at least one of 82 disease diagnoses in their UKB records, including circulatory system, renal, genitourinary, and endocrine disorders. The final set included 1051 'healthy' controls (581 female and 470 male), as shown in the left panel of Supplementary Figure 1. From the 479 post-MI subjects, 6 were excluded for incomplete image sets, 13 for missing ECG data, and 35 were excluded due to their imaging occurring before the MI event, thus retaining 425 post-MI subjects (84 female and 341 male), as shown in the right panel of Supplementary Figure 1. To select the test set used to evaluate the automated pipeline, both male and female subjects in the control population were split into deciles by their BMI. Then each decile was evenly sampled to create a test set that spanned the range of BMI for both sexes. The range of ages of the test subjects was then calculated to ensure that they represented the underlying population.

Note that the UKB protocol for imaging and ECG involved both being performed in a supine position.

**1.2 Definition of anatomical parameters.** The torso volume was calculated using PyVista [65], as the volume enclosed by the reconstructed torso surface. The cardiac volume was estimated from the extracted cardiac contours [66]. Volume of the cavity was obtained using Simpson's rule over the segmented endocardial surfaces – as described in Beetz et al. 2023 [67] this provides good estimates of cardiac volumes. Left ventricular (LV) mass was measured as the volume difference between LV epicardial and endocardial surfaces multiplied by 1.05 [68].

The x position (lateral) of the heart relative to the ECG electrodes was determined by setting the location of the most rightward electrode on the body in the torso frame as x = 0 and the most leftward as x = 1, and noting the position of the centre of the heart along this scale. y (posterior) and z (superior) positions were found similarly by setting the most frontward electrode as y = 0, backward as y = 1, downward as z = 0, and upward as z = 1.

For estimating the heart centre position, the intersection point of the LAX four-chamber plane, LAX two-chamber plane, and the middle slice of the SAX stack was identified. For estimating the heart orientation, the X-axis was first measured in the cardiac system as the intersecting line between the LAX four-chamber plane and the mid SAX slice, while the Z-axis was identified as the perpendicular line to the X-axis that passed through the LAX four-chamber plane. The Y-axis was estimated as normal to the XZ-plane. For calculating the Euler angles (α, β, γ), as shown in Supplementary Figure 3, the intersecting line, denoted as x', between the XY-plane in the cardiac system and the xy-plane in the real (torso) coordinate system was first identified. α was measured as the angle between the x-axis in the real system and the intersecting line x', essentially describing the spin of the XY-plane around the z-axis. β was measured as the angle between the Z-axis in the cardiac system and the z-axis in the real (torso) system, i.e., the verticality of the cardiac long axis. The γ angle was calculated as the angle between the X-axis in the cardiac system and the intersecting line, i.e. the tilt of the X-axis with respect to the xy-plane. The calculations of the coordinate systems and angles are presented in Supplementary Figures 2 and 3.

**1.3 Reconstruction validation.** In order to evaluate the performance of the contour extraction procedure, reconstructions of 30 subjects made using the fully automated pipeline were compared with those made using manually annotated contours. A ray tracing method [69] was applied to obtain the distance between 3D torso surfaces. For each point on one surface, the normal to the surface was found, and the nearest intersection between that line and the second surface (in either direction) was obtained. Since the normal is not guaranteed to intersect the surface or may intersect at a point on the opposing side of the torso, the nearest neighbouring point on the second surface was also found for each point. The minimum between these two distances was taken for each point as the surface-to-surface distance. The distance between each electrode location on the different torsos was also taken as a measure of similarity for the reconstructions. The mean surface-to-contour distance was calculated for each test subject to evaluate the quality of the reconstruction.

**1.4 Statistical methods.** *Choice of ECG parameters.* ECG parameters were chosen due to their availability in the UKB dataset and their relevance to diagnosis and risk stratification of MI. A range of durations, amplitudes and axis angles were chosen in order to investigate the differing effect of anatomical parameters on each category.

*Distribution of anatomical and ECG parameters for all subpopulations.* Split violin plots were created to show the distribution of each anatomical and electrophysiological parameter separately for healthy males, healthy females, post-MI males and post-MI females. Outliers were removed, as defined by values that were more than three times the inter quartile range above the third quartile or similarly below the first quartile. Mean values of all biomarkers were compared between healthy and post-MI subjects of either sex, and male and female subjects of either MI status. The Shapiro-Wilk test was used to ascertain whether both subpopulations had a normal distribution. If there was significant evidence of non-normality, the Wilcoxon test was used. The Levene test was used to determine whether the subpopulations had an equal variance. If there was sufficient evidence of nonequal variance, the Welch's t-test was used, otherwise, the standard t-test was used. For populations with statistically significant differences between them, arrows were drawn from the lower mean population to the higher mean population with stars denoting the statistical significance of the relationship. * denotes relationships with p-values between 0.05 and 0.01, ** between 0.01 and 0.001, *** between 0.001 and 0.0001, and **** less than 0.0001. The mean and standard deviation of each anatomical and ECG biomarker was recorded, alongside the p-values for differences between subpopulations.

*Age and body mass index (BMI) association.* Each anatomical parameter was separately regressed against age for the healthy male, healthy female, post-MI male, and post-MI female populations. This was repeated against BMI. This was recorded in a heat map and associated table, with the confidence intervals and two-sided p-values for the regression coefficient to be non-zero.

*Normalised correlation coefficients.* As the two measures of cardiac size (total cavity volume and LV mass) were highly colinear, one was selected for each ECG biomarker depending on the minimisation of mean Akaike information criterion (AIC) across the 12 leads. For each lead, the ECG parameter was regressed against the chosen cardiac size parameter, torso volume, the relative heart centre positions in the x, y and z directions and the three cardiac orientation parameters (spin, verticality and tilt). These anatomical parameters were chosen in order to reduce collinearity, and their variance inflation factors were all acceptably low, as shown in Supplementary Tables 2 and 3. In order to compare the regression coefficients for anatomical and ECG parameters on different scales, the regression coefficients were normalised by multiplying by the standard deviation of the anatomical parameter and then dividing by the standard deviation of the ECG biomarker, before being plotted. This was done separately for the healthy male, healthy female, post-MI male and post-MI female populations.

*Sex differences.* The contribution of each anatomical and electrophysiological parameter to the sex difference in an ECG parameter was estimated using linear regression as follows:

1. Separate the population into post-MI and healthy populations, and do the following procedure on each.
2. Regress the ECG parameter against the anatomical parameters (cavity volume/LV mass, torso volume, relative heart centre positions in the x, y, and z directions, and the three cardiac orientation parameters), and also the sex categorical parameter.
3. Calculate the difference in anatomical parameters between the male and female subpopulations.
4. Multiply the anatomical sex difference in the parameter by its regression coefficient to give its estimated contribution to the sex difference in the ECG parameter.
5. The regression coefficient for sex is the remaining sex difference when all of the chosen anatomical parameters were controlled for, so is taken as an estimate for the electrophysiological contribution to the ECG sex difference.

*MI status differences.* Similarly, the contribution of each anatomical and electrophysiological parameter to the difference between the post-MI and healthy populations was estimated using linear regression as follows:

1. Separate the population into male and female populations and do the following procedure on each.
2. Regress the ECG parameter against the anatomical parameters (cavity volume/LV mass, torso volume, relative heart centre positions in the x, y, and z directions, and the three cardiac orientation parameters), and also the healthy/post-MI categorical parameter.
3. Calculate the difference in anatomical parameters between the healthy and post-MI subpopulations.
4. Multiply the anatomical MI status difference in the parameter by its regression coefficient to give its estimated contribution to the MI status difference in the ECG parameter.
5. The regression coefficient for MI is the remaining MI status difference when all of the chosen anatomical parameters were controlled for, so is taken as an estimate for the electrophysiological contribution to the ECG MI status difference.

*Correction for anatomical contributions from demographic parameters in the healthy population.* For each subject, the difference between the population mean anatomical parameter and the subject parameter was multiplied by the regression coefficient between the ECG biomarker and that anatomical parameter. These contributions were summed to give the overall contribution of anatomy to the difference between the ECG parameter of the hypothetical 'mean subject' and that subject. Then, a linear model was made estimating this anatomical contribution from the subject's sex, height, BMI, and age. This correction was subtracted from the anatomical contribution to give a corrected anatomical contribution for each subject.

**Supplementary Appendix 2: Results**

**2.1 Anatomy-QRS duration relationship.** Mean AIC across leads, a measure of prediction error, was lower with cavity volume than LV mass (12311 and 12316 respectively). Therefore, cavity volume was chosen as the representative of cardiac size for QRSd.

**2.2 Anatomy-STj amplitude relationship.** Mean AIC across leads was lower with LV mass than cavity volume (-4841 and -4834 respectively). Therefore, LV mass was chosen as the representative of cardiac size for STj amplitude.

The analysis for sex differences in the ST amplitudes was repeated with differing measurement times to examine the sensitivity of the ECG-anatomy relationship to the method of calculating ST amplitude. STX amplitude was defined as the amplitude at time QRS offset (J point) + RR interval/16. STE amplitude was similarly the amplitude at time QRS offset + RR interval/8. Whilst the overall scale of the amplitude was higher for STX than STj, and higher still for STE, the obtained pattern of anatomical contributions to the sex difference in amplitude was conversed. As for STj, the smaller female ventricles decreased anteroseptal ST amplitude, and smaller female torso increased anteroseptal ST amplitude, for each ST calculation point.

The analysis for STj amplitude was included in the main manuscript, as this is the point at which ST-deviation should be measured in acute MI diagnosis, according to the Fourth Universal definition of Myocardial Infarction [50].

**2.3 Anatomy-TWA amplitude relationship.** Mean AIC across leads was lower with cavity volume than LV mass (-965 and -963 respectively). Therefore, cavity volume was chosen as the representative of cardiac size for TWA.

**2.4 Anatomy-axis angles relationship.** Mean AIC across R and T axes was lower with cavity volume than LV mass (14659 and 14660 respectively). Therefore, cavity volume was chosen as the representative of cardiac size for both axis angles.

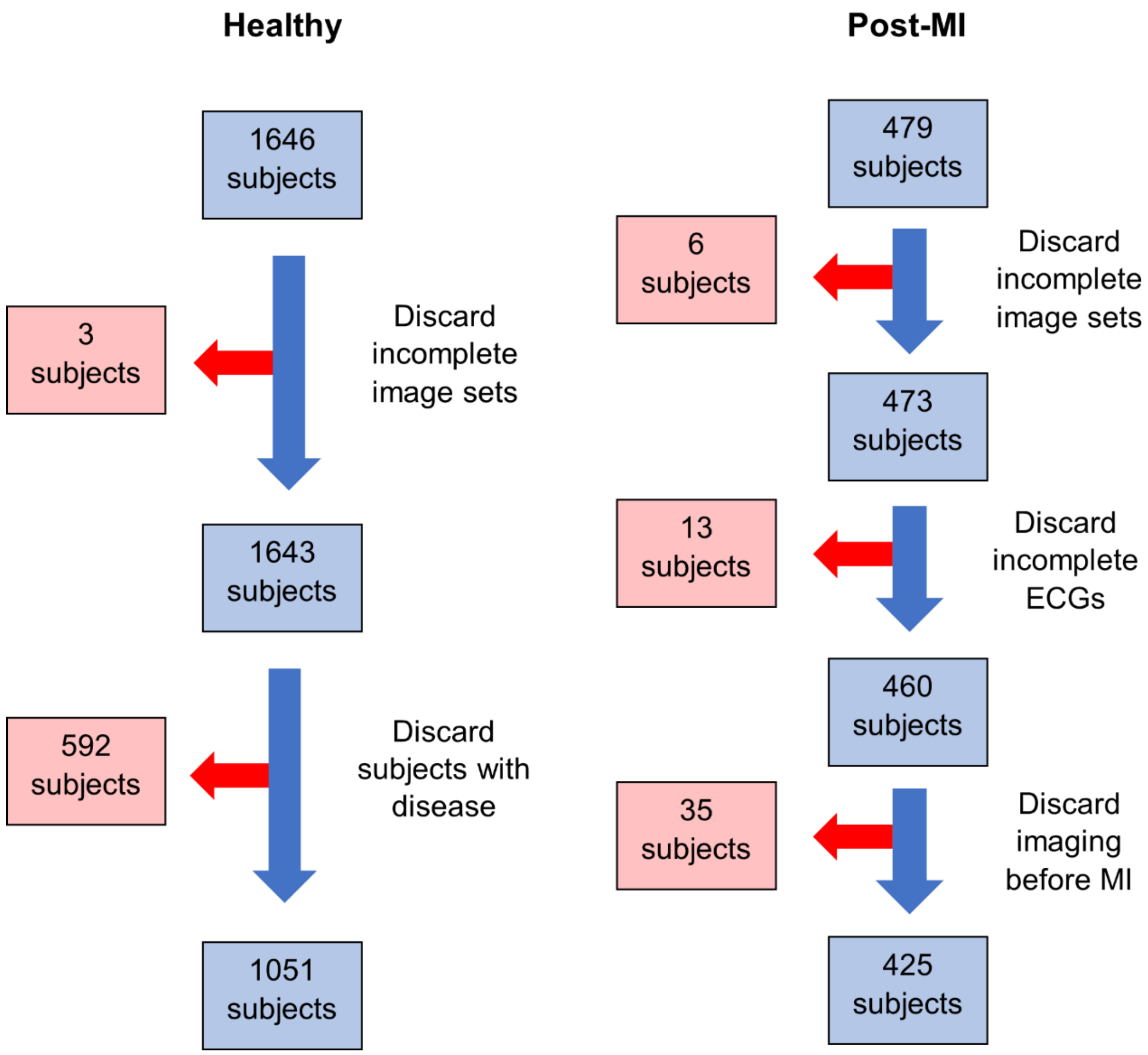

**Supplementary Figure 1**. **Exclusion pipeline for reconstruction dataset.** Both healthy and post-myocardial infarction (MI) subjects were required to have sufficient images across the view subtypes to reconstruct the torso and cardiac geometries. Subjects also were excluded if the ECG was missing or invalid (all healthy subjects had completed ECG). In the healthy dataset, subjects with at least 1 of 82 disease diagnoses were omitted, including circulatory system disorders, renal disease, genitourinary disorders, and endocrine disorders. The most common reason for exclusion was primary hypertension. Post-MI subjects were excluded if the recorded date of the MI event was after that of the imaging visit where the cardiac magnetic resonance images and ECG were taken.

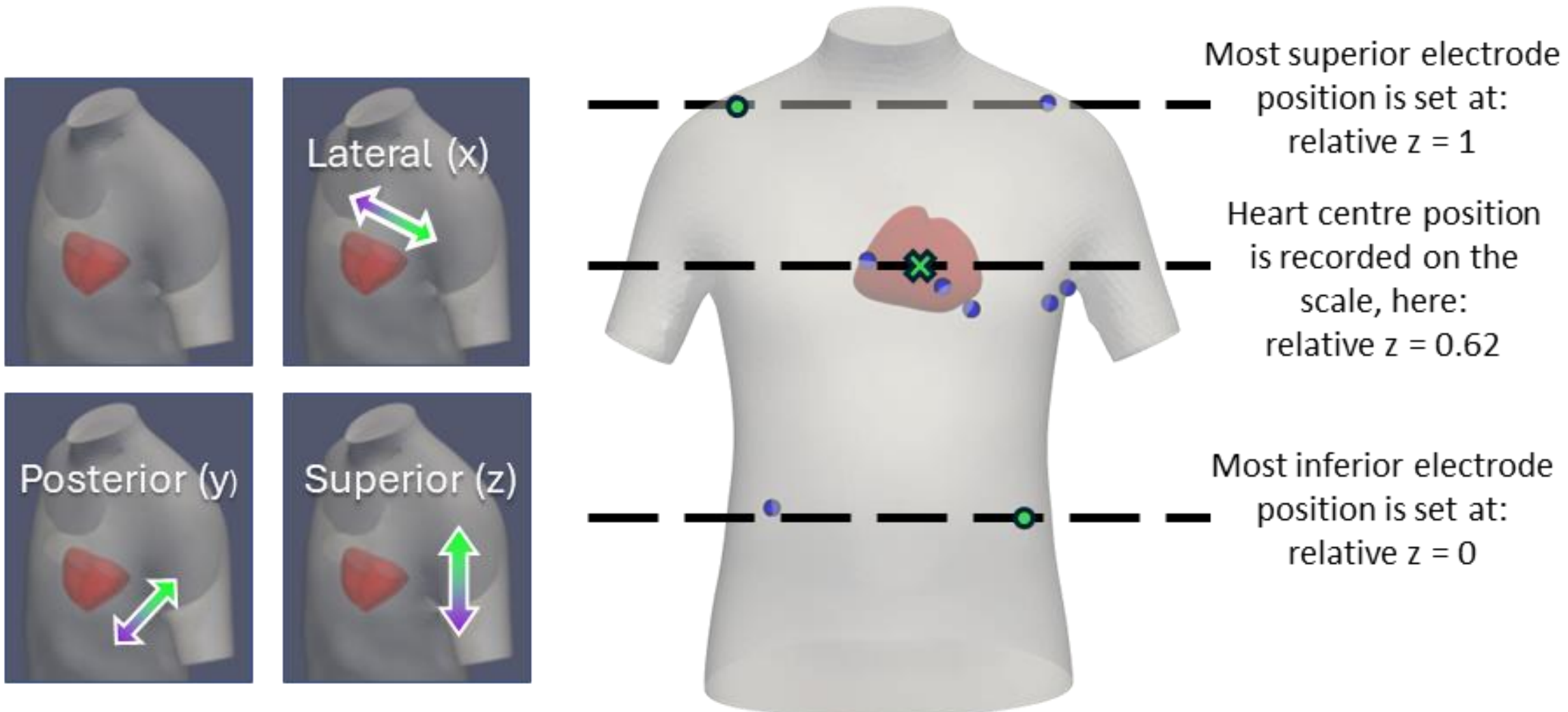

**Supplementary Figure 2. Depiction of the relative coordinate system in which the heart position was measured.** Left: three principle directions on an example torso-cardiac anatomy. Right: the calculation of the relative superior position of the cardiac centre on an example anatomy. Blue circles: ECG electrodes. Green circles: most superior and inferior electrodes. Green cross: heart centre position.

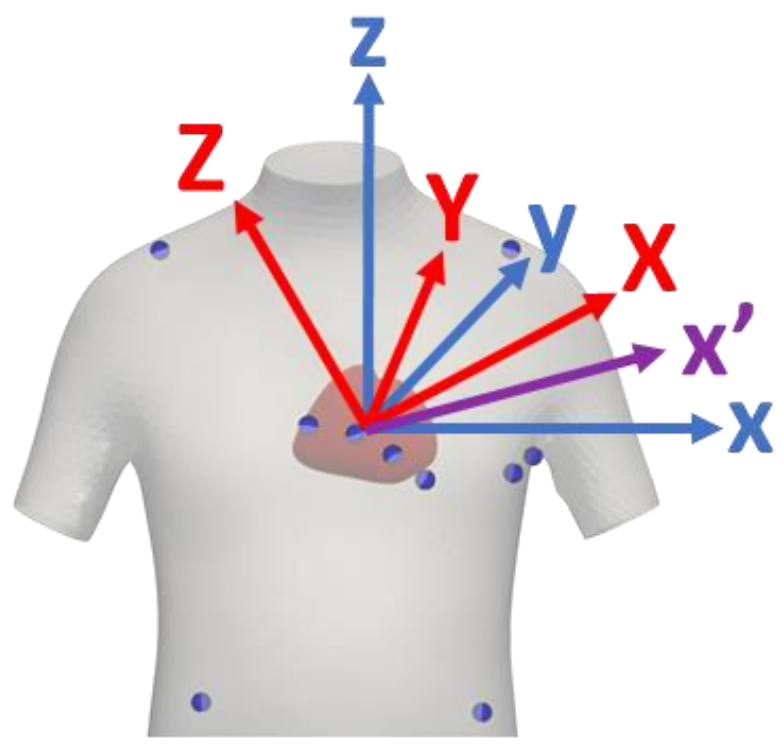

x, y, z: torso axes
X, Y, Z: cardiac axes
x': projection of cardiac X axis on torso short axis (x-y) plane

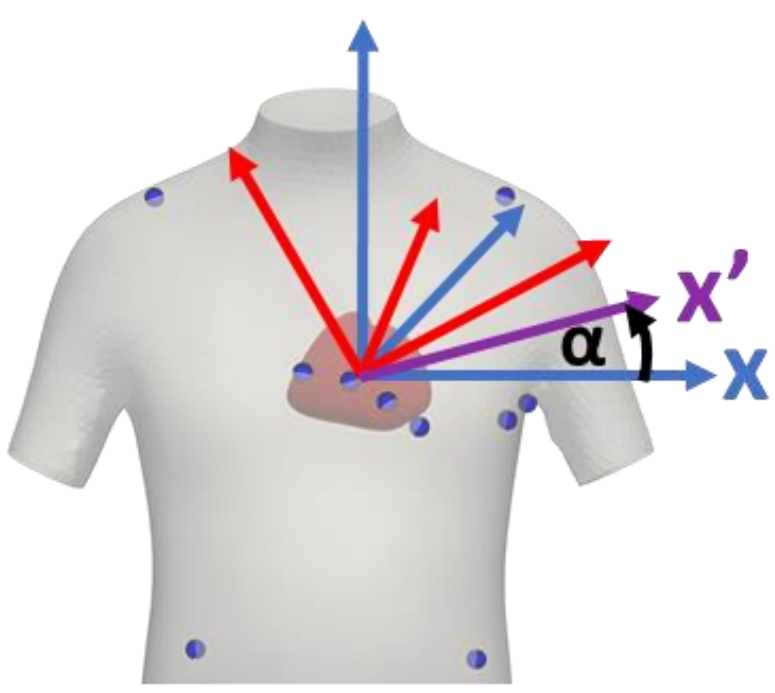

**α:** angle between **x** and **x'** on the torso short axis plane
Essentially the **spin** around the torso **z** axis

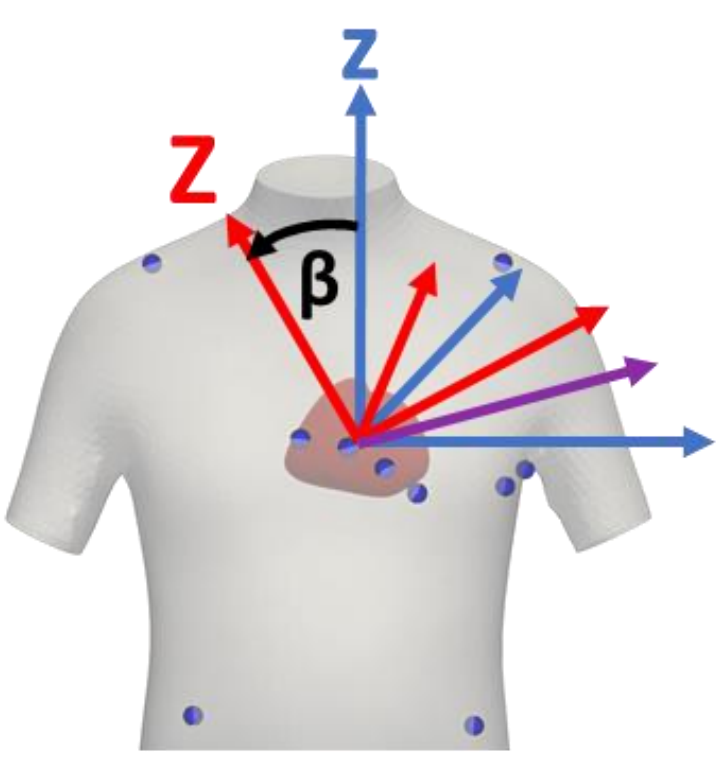

**β:** angle between cardiac long axis (**Z**) and torso long axis (**z**)
Essentially the **verticality** of the cardiac long axis

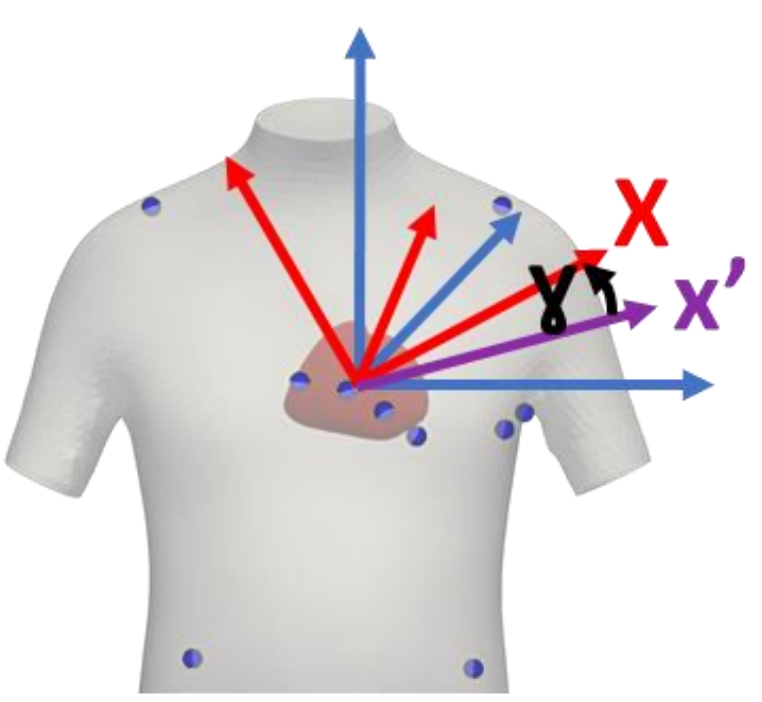

**γ:** angle between **x'** and **X**
Essentially the **tilt** of the cardiac short axis plane (**X-Y**) from the torso short axis plane (**x-y**)

**Supplementary Figure 3. Depiction of the parameters used to describe the cardiac orientation with respect to the torso.** Upper left: cardiac and torso axes, upper right: cardiac spin orientation calculation, lower left: cardiac verticality orientation calculation, lower right: cardiac tilt orientation calculation.

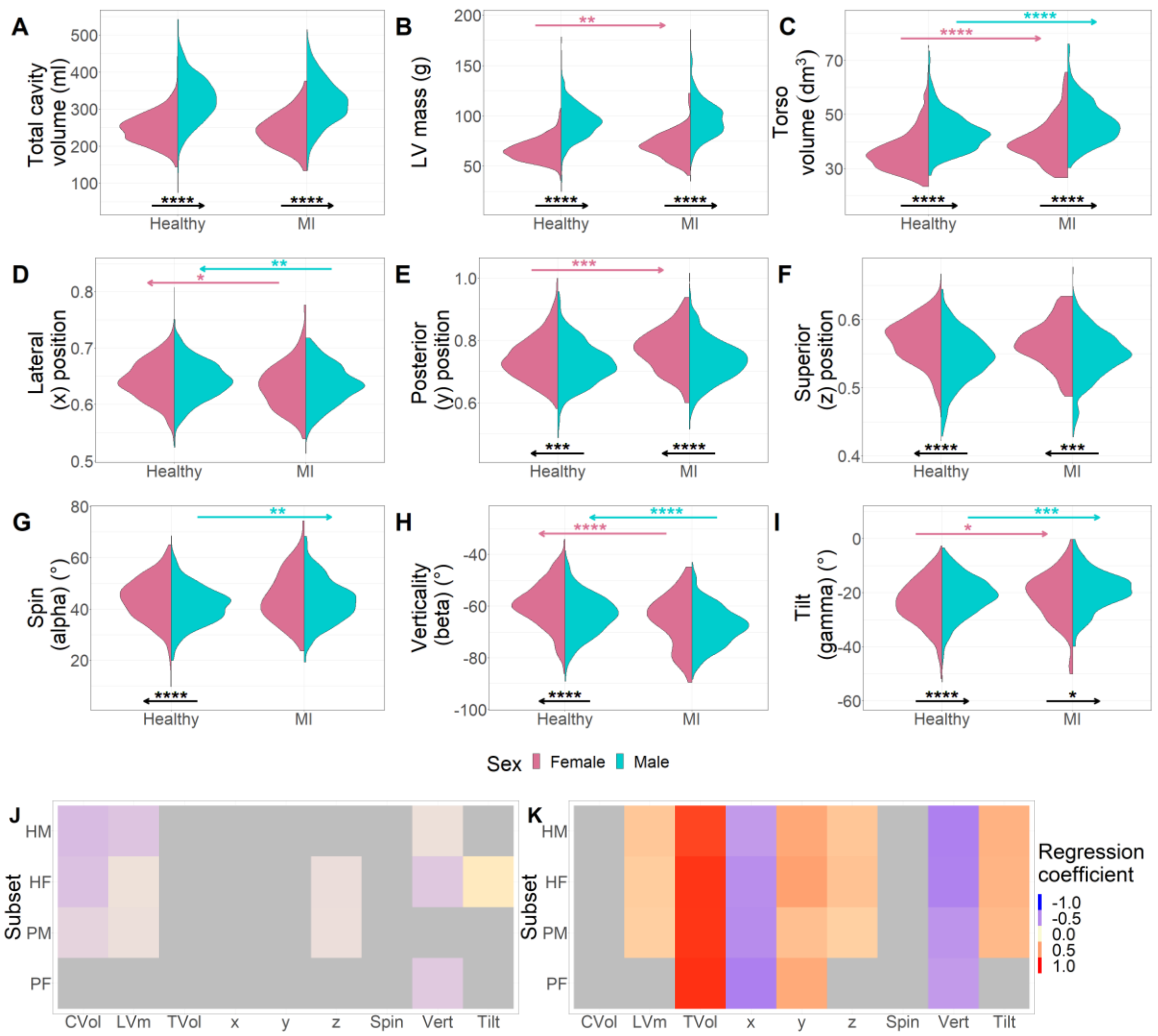

**Supplementary Figure 4. Anatomical biomarkers and their relationship with age and BMI. A-I** Distributions of anatomical biomarkers. Horizontal lines show statistically significant differences between subpopulation means with arrows pointing to the larger mean. Heat map of correlation coefficients for each geometrical parameter with age (**J**) and BMI (**K**). Correlations with p>0.05 are shaded grey. CVol: total cavity volume, LVm: left ventricular mass, TVol: torso volume, x: lateral heart centre position, y: posterior position, z: superior position, Vert: verticality of the cardiac long axis. HM: healthy male, HF: healthy female, PM: post-MI male, and PF: post-MI female.

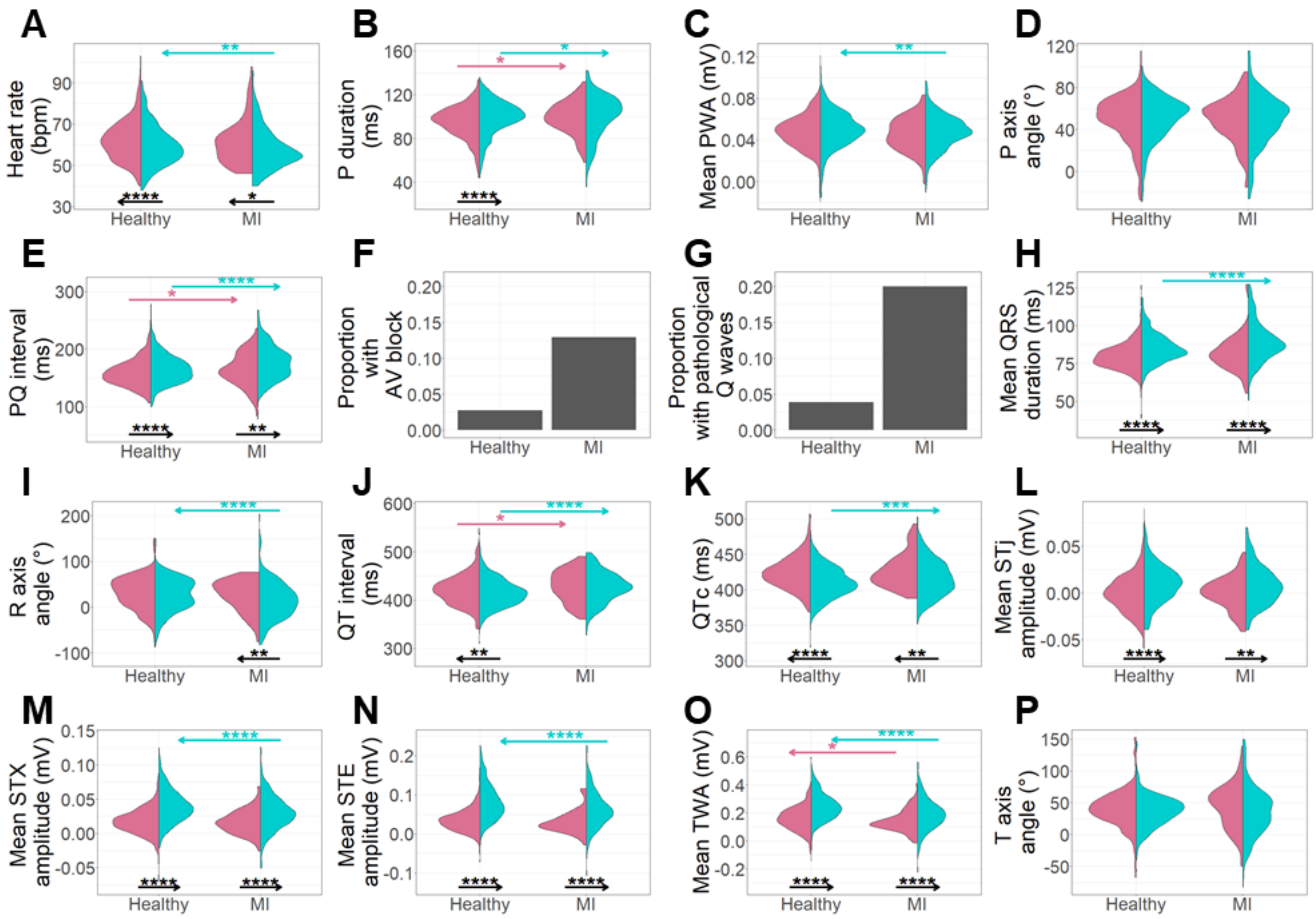

**Supplementary Figure 5. Distributions of ECG biomarkers.** Distributions of ECG biomarkers for healthy and post-MI populations and proportions of subjects with AV block and pathological Q waves. ST amplitude was measured at the J point (QRS offset), X point (QRS offset + RR interval/16) and E point (QRS offset + RR interval/8). Horizontal lines show statistically significant differences between subpopulation means with arrows pointing to the larger mean. Many distributions show features of non-normality, such as skew, heavy tails or bimodality. Post-MI subjects had longer PQ and QT intervals and P wave duration, and a lower T amplitude for both sexes. Post-MI than healthy males also had a slower heart rate, lower P amplitude, longer QRS duration, left-deviated R axis, and longer QTc. Post-MI subjects were substantially more likely to have AV block and pathological Q waves. The T axis showed a higher variability in post-MI individuals for both sexes. AV: atrioventricular, QTc: corrected QT interval.

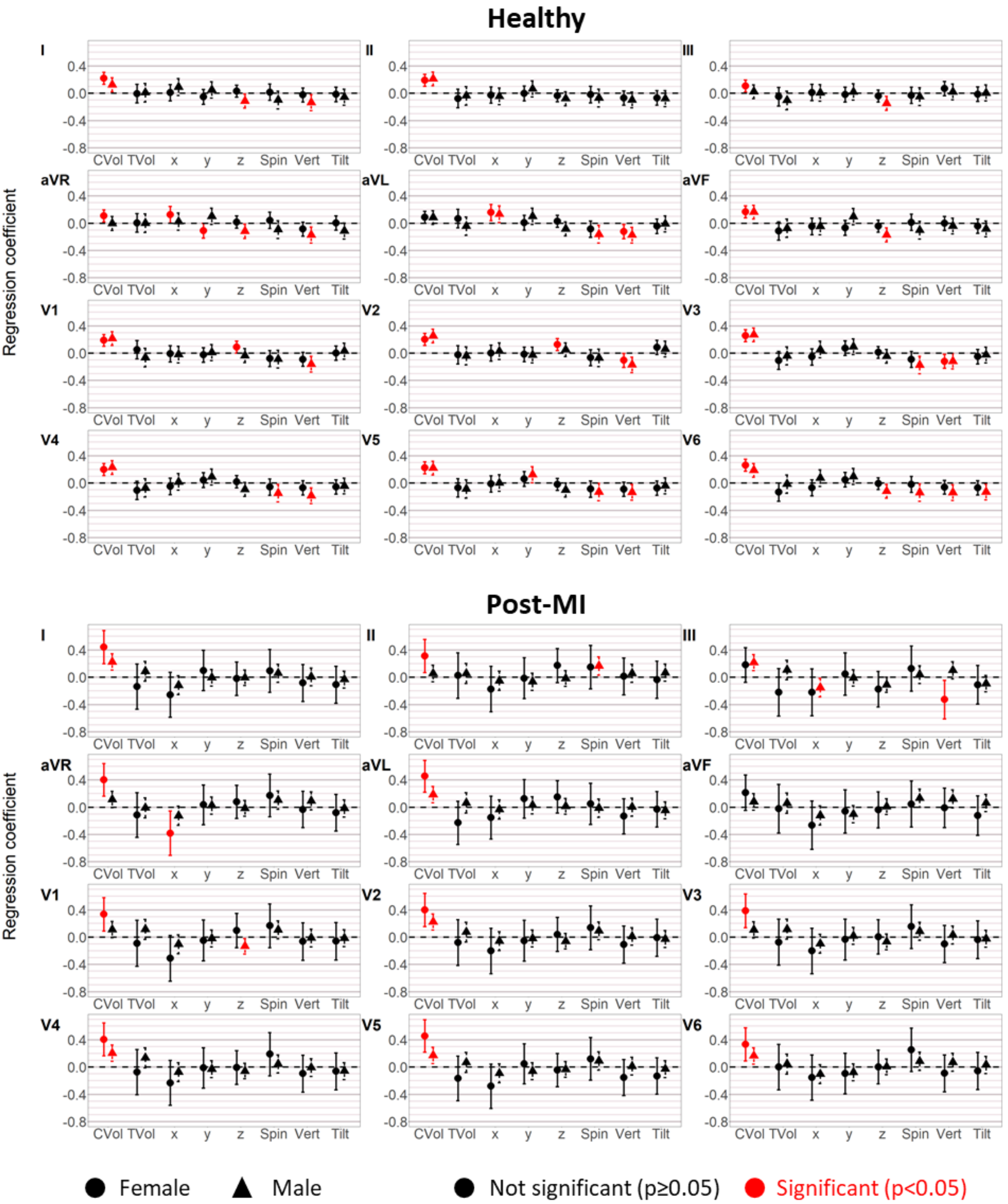

**Supplementary Figure 6. QRS duration-anatomy relationship.** Normalised regression coefficients with 95% confidence intervals for QRS duration (QRSd) in all ECG leads for healthy (upper) and post-MI (lower) subjects. CVol: total cavity volume, TVol: torso volume, x, y, z: x (medial), y (posterior), z (superior) coordinate of the heart centre relative to the electrodes, Vert: verticality of the cardiac long axis. Bars shown in red represent regression coefficients that are significantly different from 0, at a significance level of 0.05. QRSd was shortened by a more vertical cardiac orientation (less negative β) for at least one sex in a majority of leads. Increasing cavity volume prolonged the QRSd. QRSd was largely unaffected by adjustments in the cardiac centre location. Unlike healthy subjects, for post-MI individuals QRSd was largely unaffected by cardiac orientation. QRSd again usually did not significantly vary with cardiac location.

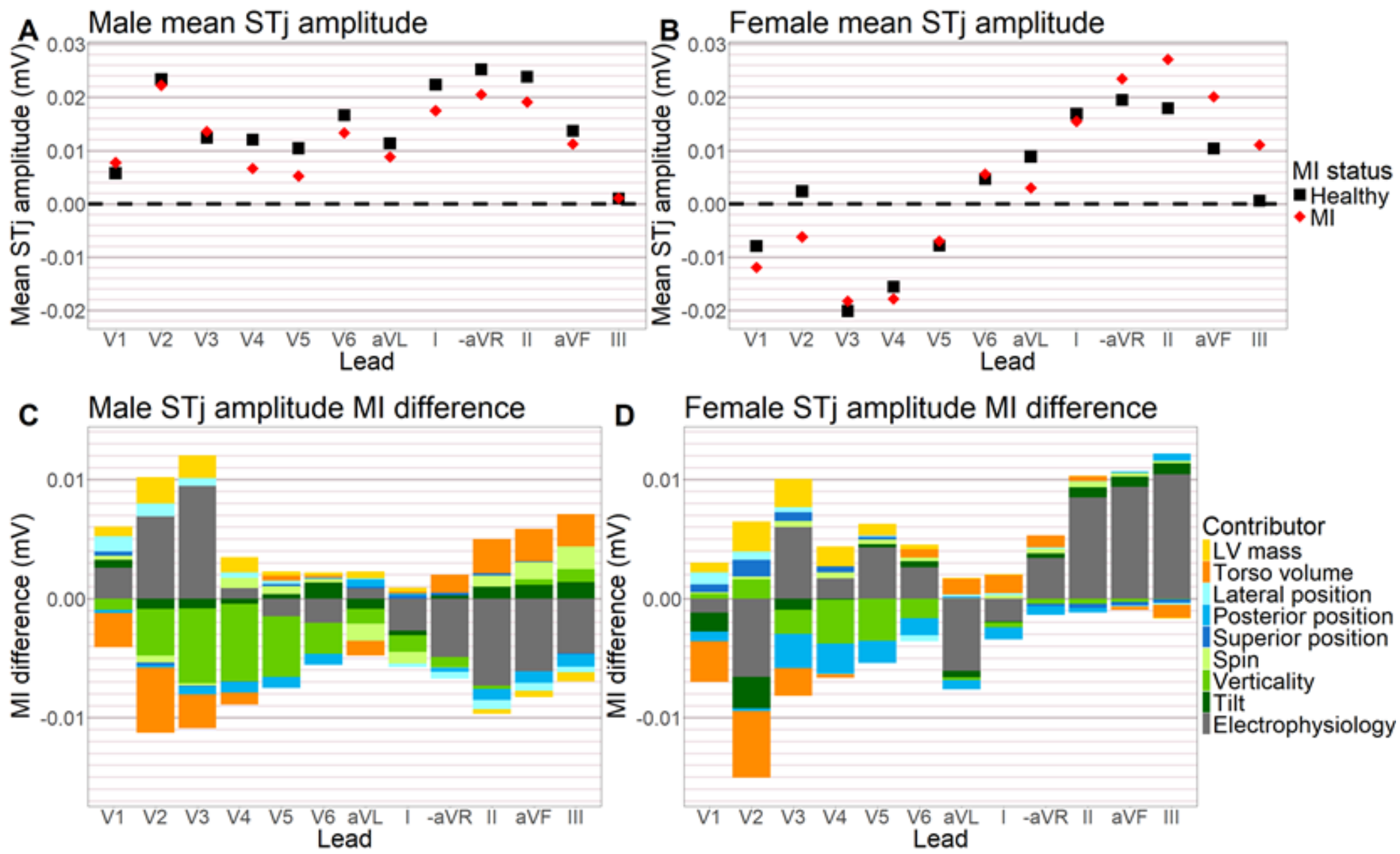

**Supplementary Figure 7. Difference between healthy and post-MI STj amplitude, separated by sex. A**, **B:** mean STj amplitudes for each ECG lead in men and women respectively for healthy (black squares) and post-MI (red diamonds). **C, D:** contribution of anatomical parameters and electrophysiology to differences in STj amplitudes between healthy and post-MI subjects for male and female subjects (calculated by multiplying the regression coefficient for each factor by its mean difference between healthy and post-MI populations).

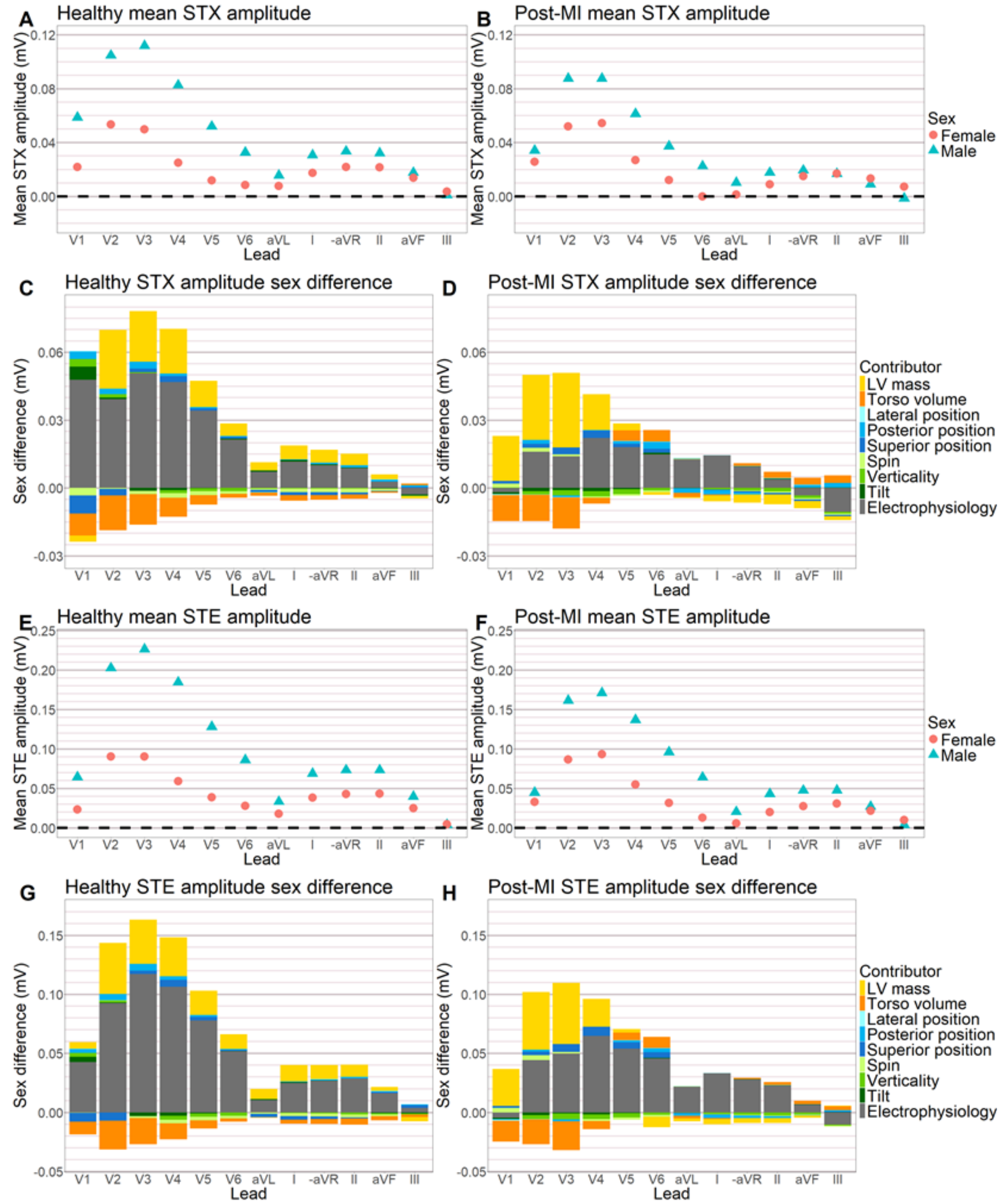

**Supplementary Figure 8. Sex differences in ST amplitudes for differing measurement points in healthy and post-MI subjects. A, B:** mean STX amplitude (measured at QRS offset + RR interval/16) for each ECG lead in healthy and post-MI subjects respectively with women (red circles) and men (cyan triangles). **C, D:** contribution of anatomical parameters and electrophysiology to sex differences in STX amplitude for healthy and post-MI subjects (calculated by multiplying the regression coefficient for each factor by its mean difference between male and female populations). **E, F:** mean STE amplitude (measured at QRS offset + RR interval/8) for each ECG lead in healthy and post-MI subjects respectively with women (red circles) and men (cyan triangles). **G, H:** contribution of anatomical parameters and electrophysiology to sex differences in STE amplitude for healthy and post-MI subjects (calculated by multiplying the regression coefficient for each factor by its mean difference between male and female populations).

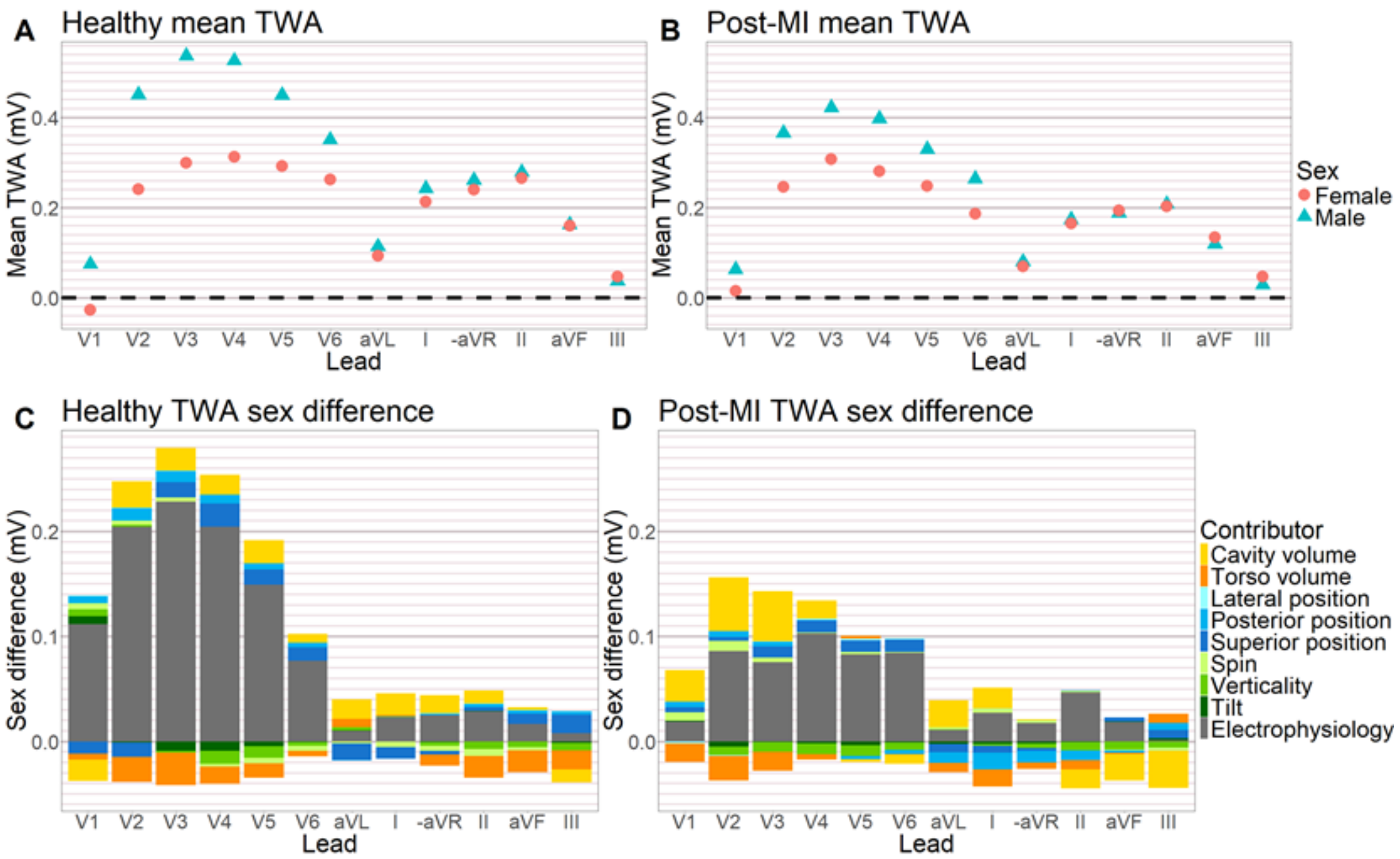

**Supplementary Figure 9. Sex differences in T wave amplitude in healthy and post-MI subjects. A, B:** mean T wave amplitude (TWA) for each ECG lead in healthy and post-MI subjects respectively with women (red circles) and men (cyan triangles). **C,D:** contribution of anatomical parameters and electrophysiology to sex differences in TWA for healthy and post-MI subjects (calculated by multiplying the regression coefficient for each factor by its mean difference between male and female populations).

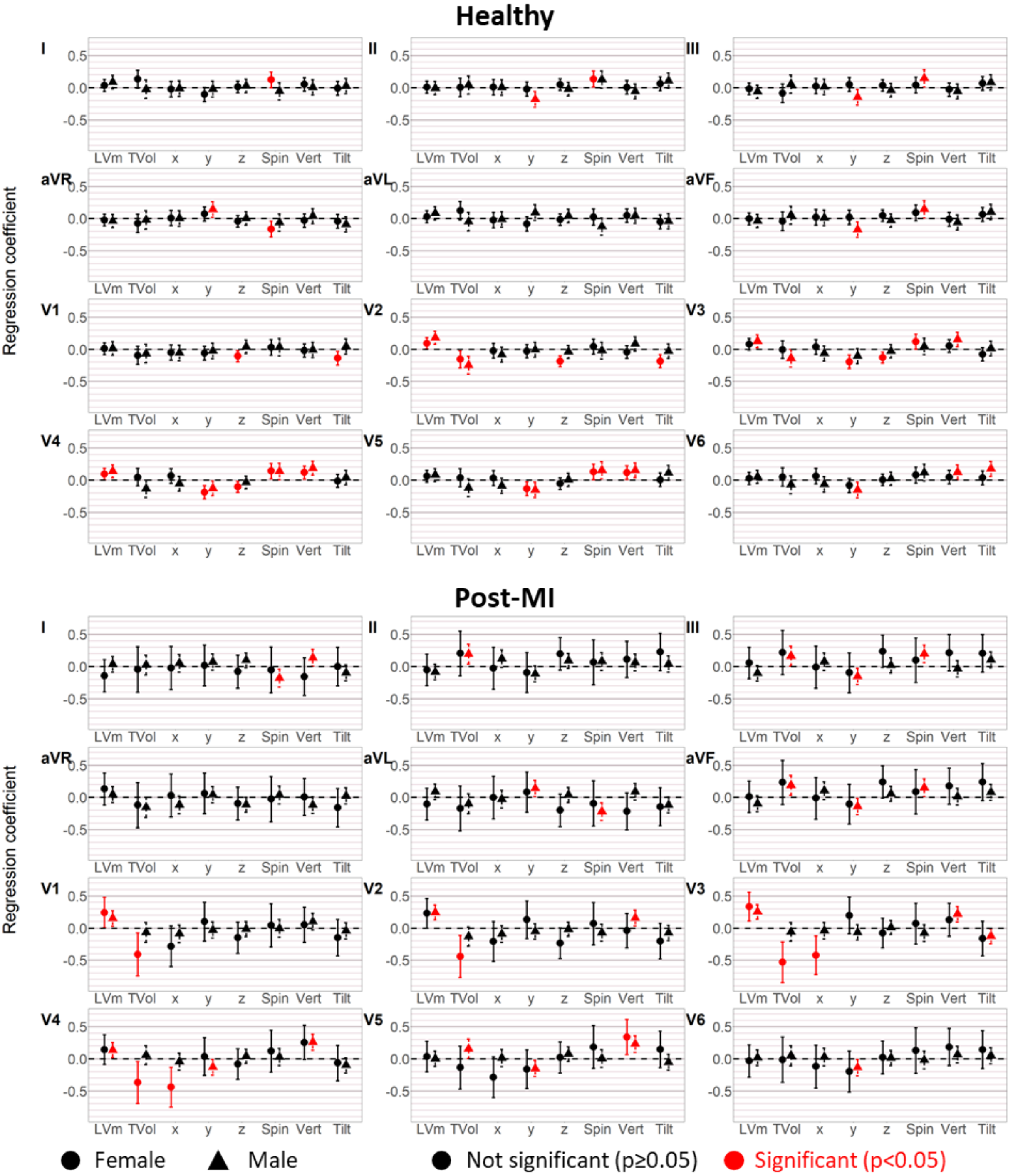

**Supplementary Figure 10. STj amplitude-anatomy relationship.** Normalised regression coefficients with 95% confidence intervals for STj amplitude in all ECG leads for healthy (upper) and post-MI (lower) subjects. LVm: left ventricular mass, TVol: torso volume, x, y, z: x (medial), y (posterior), z (superior) coordinate of the heart centre relative to the electrodes, Vert: verticality of the cardiac long axis. Bars shown in red represent regression coefficients that are significantly different from 0, at a significance level of 0.05. Significant positional correlations generally followed the pattern that the closer the heart centre was to the position of the exploring electrode, the higher the STj amplitude. Increases in left ventricular mass were associated with increased STj amplitude in some of the precordial leads.

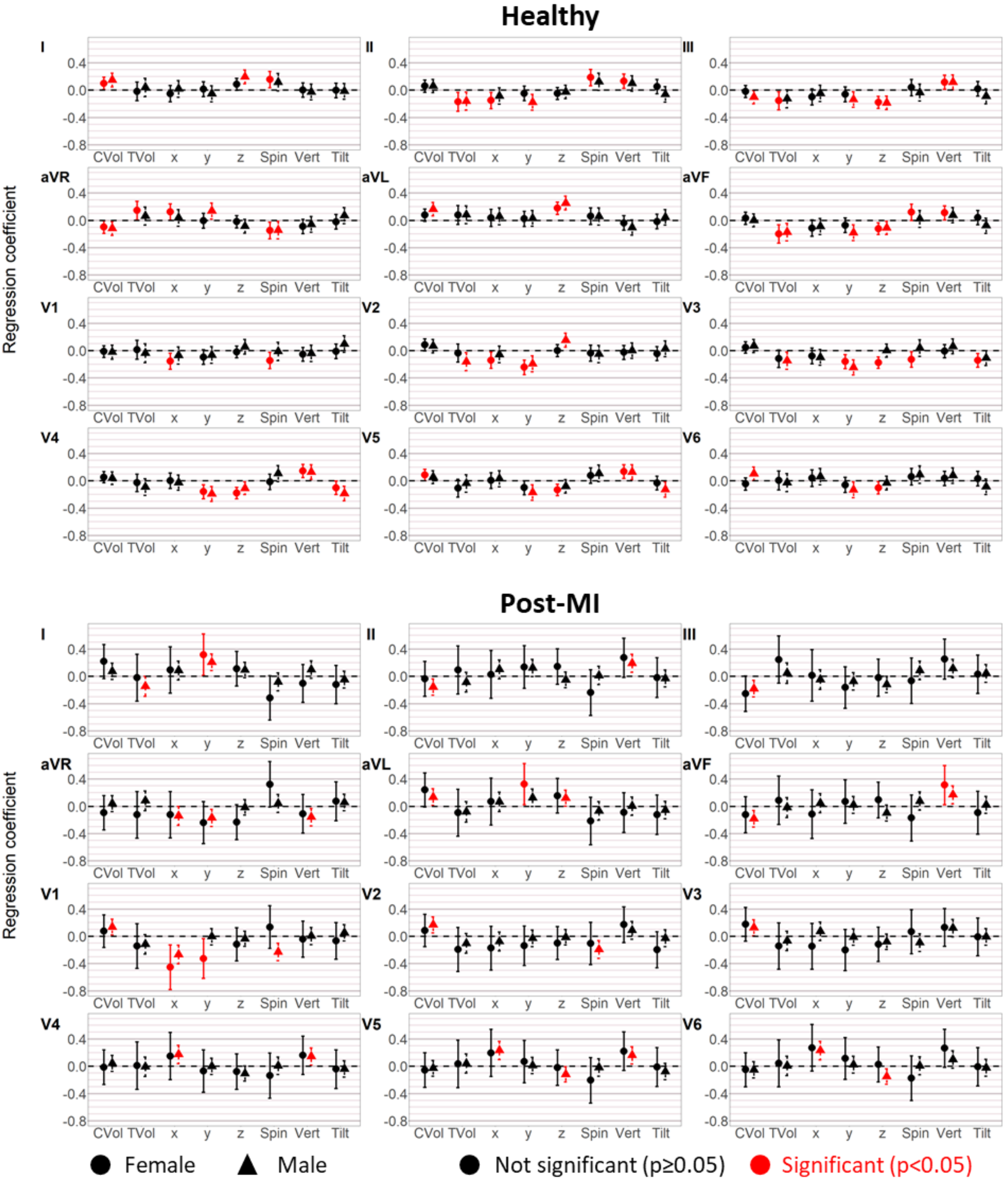

**Supplementary Figure 11. T wave amplitude-anatomy relationship.** Normalised regression coefficients with 95% confidence intervals for T wave amplitude (TWA) in all ECG leads for healthy (upper) and post-MI (lower) subjects. CVol: total cavity volume, TVol: torso volume, x, y, z: x (medial), y (posterior), z (superior) coordinate of the heart centre relative to the electrodes, Vert: verticality of the cardiac long axis. Bars shown in red represent regression coefficients that are significantly different from 0, at a significance level of 0.05. TWA in many leads increases as the heart is oriented more vertically (less negative β). Significant positional correlations generally followed the pattern that the closer the heart centre was to the position of the exploring electrode, the higher the TWA.

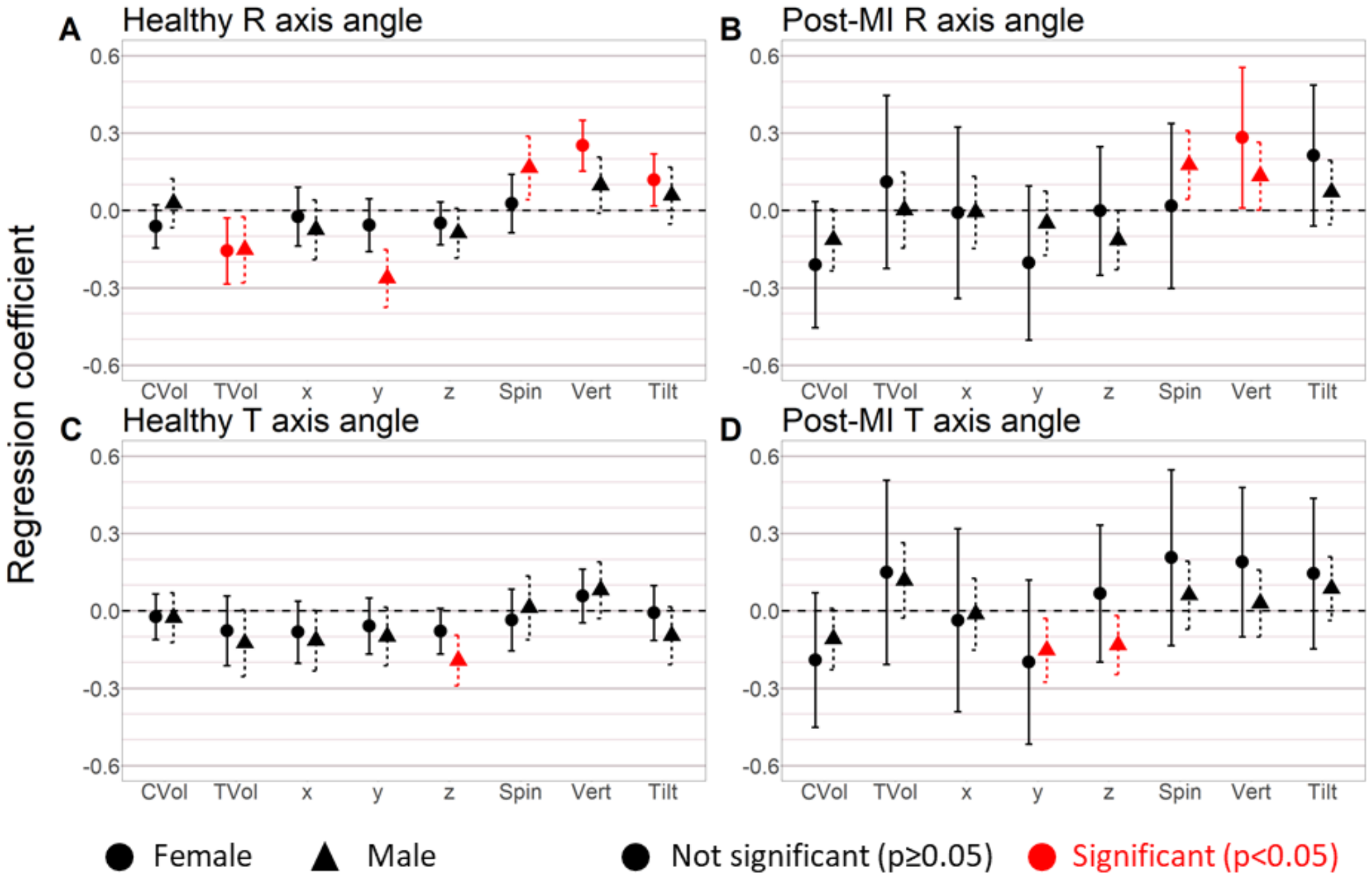

**Supplementary Figure 12. Axis angles-anatomy relationship.** Normalised regression coefficients with 95% confidence intervals for R axis (top) and T axis (bottom), in healthy (left) and post-MI (right) subjects. For healthy but not post-MI subjects, R axis was negatively associated with torso volume. R axis was significantly affected by cardiac orientation, but not T axis. CVol: total cavity volume, TVol: torso volume, x, y, z: x (medial), y (posterior), z (superior) coordinate of the heart centre relative to the electrodes, Vert: verticality of the cardiac long axis. Bars shown in red represent regression coefficients significantly different from 0, at a significance level of 0.05.

**Supplementary Table 1. Disease and treatment characteristics of 425 post-MI subjects from the UK Biobank cohort.**

| **Characteristics** | **Male (N = 341)** | **Female (N = 84)** | **p-value** |
|---|---|---|---|
| **Disease prevalence (%)** | | | |
| Primary (essential) hypertension | 241 (70.7) | 50 (59.5) | 0.07 |
| Atrial fibrillation or flutter | 60 (17.6) | 10 (11.9) | 0.3 |
| Heart failure | 49 (14.4) | 10 (11.9) | 0.1 |
| Renal failure | 34 (10.0) | 11 (13.1) | 0.5 |
| Diabetes | 29 (8.5) | 6 (7.1) | 0.9 |
| **Surgical treatment (%)** | | | |
| Replacement of coronary artery | 52 (15.2) | 1 (1.2) | 0.0009 |
| Bypass of coronary artery | 57 (16.7) | 2 (2.4) | 0.001 |
| Balloon angioplasty of coronary artery without stent | 41 (12.0) | 5 (6.0) | 0.2 |
| Balloon angioplasty of coronary artery with stent | 135 (39.6) | 21 (25.0) | 0.02 |

Categorical variables are shown as number of subjects (percentage in bracket). $\chi^2$-tests were performed to assess the proportions of subjects with each disease or treatment. The p-value refers to the null hypothesis that the proportion in a category is equal in the male and female subpopulations. Surgical treatment information was sourced from hospital records, whereas disease information was obtained from multiple sources including hospital records, primary care and self-report.

**Supplementary Table 2. Variance inflation factors for linear regression models with total cavity volume.**

| | Healthy male | Healthy female | MI male | MI female | Healthy (both sexes) | MI (both sexes) | Male | Female |
|---|---|---|---|---|---|---|---|---|
| **Total cavity volume** | 1.3 | 1.2 | 1.3 | 1.4 | 1.9 | 1.6 | 1.3 | 1.2 |
| **Torso volume** | 2.3 | 2.8 | 1.9 | 2.6 | 3.1 | 2.2 | 2.2 | 2.8 |
| **Lateral position** | 1.9 | 2.2 | 1.8 | 2.5 | 2.1 | 1.8 | 1.8 | 2.3 |
| **Posterior position** | 1.8 | 1.8 | 1.4 | 2.0 | 1.8 | 1.5 | 1.6 | 1.9 |
| **Superior position** | 1.3 | 1.2 | 1.2 | 1.4 | 1.4 | 1.2 | 1.2 | 1.2 |
| **Spin** | 2.1 | 2.2 | 1.6 | 2.3 | 2.2 | 1.7 | 1.8 | 2.2 |
| **Verticality** | 1.7 | 1.7 | 1.5 | 1.7 | 1.7 | 1.6 | 1.7 | 1.7 |
| **Tilt** | 1.7 | 1.7 | 1.4 | 1.7 | 1.8 | 1.4 | 1.6 | 1.7 |
| **Sex** | N/A | N/A | N/A | N/A | 2.2 | 1.6 | N/A | N/A |
| **MI status** | N/A | N/A | N/A | N/A | N/A | N/A | 1.1 | 1.1 |

Calculated variance inflation factors, reflecting the collinearity of anatomical and demographic biomarkers for all parameter sets used. Linear regression models were obtained using the sum of all anatomical (and no demographic) parameters with total cavity volume as the cardiac size metric for healthy male/female and MI male/female populations, the sum of all anatomical parameters plus the categorical sex parameter for the mixed sex healthy/MI populations, and the sum of all anatomical parameters plus the categorical MI status parameter for the male/female populations. All parameter combinations show low degrees of collinearity, thus enabling simple linear regression.

**Supplementary Table 3. Variance inflation factors for linear regression models with left ventricular mass.**

| | Healthy male | Healthy female | MI male | MI female | Healthy (both sexes) | MI (both sexes) | Male | Female |
|---|---|---|---|---|---|---|---|---|
| **LV mass** | 1.3 | 1.3 | 1.3 | 1.3 | 2.5 | 1.7 | 1.3 | 1.3 |
| **Torso volume** | 2.5 | 3.1 | 2.2 | 2.5 | 3.3 | 2.4 | 2.4 | 3.0 |
| **Lateral position** | 1.8 | 2.2 | 1.7 | 2.2 | 2.0 | 1.7 | 1.7 | 2.2 |
| **Posterior position** | 1.8 | 1.8 | 1.4 | 2.0 | 1.8 | 1.6 | 1.6 | 1.9 |
| **Superior position** | 1.3 | 1.2 | 1.2 | 1.3 | 1.4 | 1.3 | 1.3 | 1.2 |
| **Spin** | 2.2 | 2.3 | 1.7 | 2.5 | 2.3 | 1.8 | 1.9 | 2.3 |
| **Verticality** | 1.7 | 1.7 | 1.5 | 1.7 | 1.7 | 1.5 | 1.7 | 1.7 |
| **Tilt** | 1.7 | 1.7 | 1.4 | 1.8 | 1.8 | 1.4 | 1.6 | 1.7 |
| **Sex** | N/A | N/A | N/A | N/A | 2.4 | 1.6 | N/A | N/A |
| **MI status** | N/A | N/A | N/A | N/A | N/A | N/A | 1.1 | 1.1 |

Calculated variance inflation factors, reflecting the collinearity of anatomical and demographic biomarkers for all parameter sets used. Linear regression models were obtained using the sum of all anatomical (and no demographic) parameters with left ventricular mass as the cardiac size metric for healthy male/female and MI male/female populations, the sum of all anatomical parameters plus the categorical sex parameter for the mixed sex healthy/MI populations, and the sum of all anatomical parameters plus the categorical MI status parameter for the male/female populations. All parameter combinations show low degrees of collinearity, thus enabling simple linear regression. LV: left ventricular.

**Supplementary Table 4. Mean and standard deviation values of anatomical and ECG biomarkers in each subpopulation.**

| Biomarker | Healthy | | | MI | | | Male | Female |
|---|---|---|---|---|---|---|---|---|
| | Male N=470 | Female N=581 | p-value | Male N=341 | Female N=84 | p-value | p-value | p-value |
| Total cavity volume (ml) | 326 (62) | 249 (42) | < 0.0001 | 320 (63) | 245 (47) | < 0.0001 | 0.09 | 0.4 |
| LV mass (g) | 94.3 (17.0) | 66.7 (11.0) | < 0.0001 | 97.3 (19.3) | 70.9 (13.3) | < 0.0001 | 0.06 | 0.003 |
| Torso volume ($dm^3$) | 43.4 (7.5) | 36.5 (7.6) | < 0.0001 | 47.0 (7.7) | 40.3 (8.0) | < 0.0001 | < 0.0001 | < 0.0001 |
| Lateral (x) position | 0.642 (0.034) | 0.643 (0.036) | 0.7 | 0.635 (0.035) | 0.634 (0.041) | 0.8 | 0.002 | 0.02 |
| Posterior (y) position | 0.727 (0.069) | 0.743 (0.071) | 0.0006 | 0.736 (0.070) | 0.770 (0.069) | < 0.0001 | 0.06 | 0.0004 |
| Superior (z) position | 0.544 (0.037) | 0.569 (0.033) | < 0.0001 | 0.549 (0.036) | 0.565 (0.034) | 0.0002 | 0.09 | 0.2 |
| Spin α (°) | 41.4 (7.6) | 44.0 (8.4) | < 0.0001 | 43.0 (8.4) | 44.8 (9.9) | 0.1 | 0.005 | 0.5 |
| Verticality β (°) | -63.4 (7.9) | -59.5 (9.0) | < 0.0001 | -67.7 (7.6) | -65.4 (10.0) | 0.05 | < 0.0001 | < 0.0001 |
| Tilt γ (°) | -21.2 (7.5) | -23.6 (8.4) | < 0.0001 | -19.2 (7.1) | -21.5 (9.1) | 0.03 | 0.0002 | 0.02 |
| Heart rate (I) (bmp) | 59.9 (10.4) | 62.4 (10.7) | < 0.0001 | 58.5 (11.2) | 61.2 (10.7) | 0.02 | 0.007 | 0.1 |
| P duration (I) (ms) | 98.7 (15.9) | 95.1 (14.7) | < 0.0001 | 100.3 (19.1) | 98.9 (14.8) | 0.2 | 0.04 | 0.03 |
| Mean P amplitude (mV) | 0.0755 (0.0260) | 0.0747 (0.0384) | 0.3 | 0.0707 (0.0186) | 0.0708 (0.0170) | 0.8 | 0.005 | 0.3 |
| P axis (°) | 50.9 (22.1) | 49.6 (22.4) | 0.5 | 48.6 (25.6) | 49.1 (21.6) | 0.9 | 0.2 | 0.6 |
| PQ interval (I) (ms) | 165 (25) | 158 (23) | < 0.0001 | 178 (32) | 164 (28) | 0.001 | < 0.0001 | 0.02 |
| Mean QRS duration (ms) | 85.8 (10.8) | 79.8 (9.1) | < 0.0001 | 90.7 (15.1) | 82.5 (14.5) | < 0.0001 | < 0.0001 | 0.1 |
| R axis (°) | 27.6 (35.9) | 31.2 (34.9) | 0.1 | 15.6 (41.5) | 25.8 (34.1) | 0.003 | < 0.0001 | 0.3 |
| QT interval (I) (ms) | 415 (31) | 421 (33) | 0.003 | 428 (31) | 429 (32) | 0.8 | < 0.0001 | 0.02 |
| QTc (I) (ms) | 411 (24) | 426 (29) | < 0.0001 | 418 (26) | 429 (27) | 0.002 | 0.0009 | 0.6 |
| Mean STj amplitude (mV) | 0.0106 (0.0225) | -0.0008 (0.0190) | < 0.0001 | 0.0088 (0.0231) | -0.0002 (0.0176) | 0.002 | 0.09 | 0.8 |
| Mean STX amplitude (mV) | 0.0422 (0.0510) | 0.0178 (0.0179) | <0.0001 | 0.0303 (0.0253) | 0.0170 (0.0175) | <0.0001 | <0.0001 | 0.6 |
| Mean STE amplitude (mV) | 0.0866 (0.0610) | 0.0347 (0.0283) | <0.0001 | 0.0641 (0.0482) | 0.0311 (0.0288) | <0.0001 | <0.0001 | 0.1 |
| Mean T amplitude (mV) | 0.248 (0.109) | 0.160 (0.084) | < 0.0001 | 0.189 (0.108) | 0.144 (0.077) | < 0.0001 | < 0.0001 | 0.03 |
| T axis (°) | 38.9 (30.4) | 41.0 (31.6) | 0.08 | 39.2 (43.8) | 45.6 (37.6) | 0.06 | 0.7 | 0.1 |

p-values are measured based on appropriate two-sided test (as discussed in statistical methods) for: first two columns, the mean male and female values being different and last two columns, the mean healthy and post-MI values being different. p-values are shaded red if below 0.05, i.e., statistically significant. LV: left ventricular. Post-MI hearts were found to be placed in a more medial position (lower x) and less vertical orientation (more negative β) than those of healthy subjects. For women, the cardiac position was also found to be more posterior (larger y) in post-MI than healthy subjects. Torso volume was considerably larger in post-MI subjects. Women had a more posterior (larger y) and superior (larger z) cardiac position than men and their orientation was more vertical (less negative β). Cavity volume, LV mass and torso volume were substantially lower in women than men. ST amplitude was measured at the J point (QRS offset), X point (QRS offset + RR interval/16) and E point (QRS offset + RR interval/8). Post-MI subjects had longer P durations and PQ and QT intervals, and a lower T wave amplitude for both sexes. Post-MI men also had a slower heart rate, lower P wave amplitude, longer QRS duration, left-deviated R axis, and longer QTc than healthy men. AV: atrioventricular, QTc: corrected QT interval.

**Supplementary Table 5. Associations between age and BMI and anatomical biomarkers.**

| ***Age*** | | | | |
|---|---|---|---|---|
| **Biomarker** | Healthy | | MI | |
| | Male N=470 | Female N=581 | Male N=341 | Female N=84 |
| Total cavity volume (ml) | -0.277 [-0.364, -0.189], p < 0.0001 | -0.251 [-0.330, -0.172], p < 0.0001 | -0.172 [-0.277, -0.067], p = 0.001 | -0.042 [-0.261, 0.178], p = 0.7 |
| LV mass (g) | -0.238 [-0.326, -0.150], p < 0.0001 | -0.107 [-0.188, -0.026], p = 0.01 | -0.117 [-0.223, -0.011], p = 0.03 | -0.064 [-0.283, 0.155], p = 0.6 |
| Torso volume ($dm^3$) | -0.021 [-0.111, 0.070], p = 0.7 | -0.014 [-0.096, 0.068], p = 0.7 | -0.024 [-0.131, 0.083], p = 0.7 | -0.195 [-0.410, 0.020], p = 0.08 |
| Lateral (x) position | 0.062 [-0.029, 0.153], p = 0.2 | 0.032 [-0.050, 0.113], p = 0.4 | -0.056 [-0.163, 0.051], p = 0.3 | 0.039 [-0.181, 0.258], p = 0.7 |
| Posterior (y) position | 0.063 [-0.027, 0.154], p = 0.2 | 0.039 [-0.042, 0.121], p = 0.3 | -0.041 [-0.148, 0.066], p = 0.4 | -0.201 [-0.416, 0.014], p = 0.07 |
| Superior (z) position | -0.070 [-0.161, 0.021], p = 0.1 | -0.127 [-0.208, -0.046], p = 0.002 | -0.123 [-0.229, -0.017], p = 0.02 | 0.057 [-0.162, 0.276], p = 0.6 |
| Spin α (°) | -0.089 [-0.180, 0.001], p = 0.05 | 0.027 [-0.054, 0.109], p = 0.5 | -0.103 [-0.210, 0.003], p = 0.06 | 0.060 [-0.160, 0.279], p = 0.6 |
| Verticality β (°) | -0.113 [-0.203, -0.022], p = 0.01 | -0.218 [-0.298, -0.139], p < 0.0001 | -0.100 [-0.206, 0.007], p = 0.07 | -0.215 [-0.429, 0.000], p = 0.05 |
| Tilt γ (°) | 0.068 [-0.023, 0.158], p = 0.1 | 0.088 [0.007, 0.170], p = 0.03 | 0.052 [-0.055, 0.158], p = 0.3 | -0.080 [-0.299, 0.139], p = 0.5 |
| ***BMI*** | | | | |
| **Biomarker** | Healthy | | MI | |
| | Male N=470 | Female N=581 | Male N=341 | Female N=84 |
| Total cavity volume (ml) | 0.000 [-0.093, 0.093], p = 1 | 0.065 [-0.017, 0.148], p = 0.1 | -0.047 [-0.155, 0.062], p = 0.4 | 0.028 [-0.203, 0.259], p = 0.8 |
| LV mass (g) | 0.277 [0.188, 0.367], p < 0.0001 | 0.247 [0.167, 0.327], p < 0.0001 | 0.235 [0.129, 0.340], p < 0.0001 | 0.198 [-0.029, 0.425], p = 0.09 |
| Torso volume ($dm^3$) | 0.878 [0.846, 0.909], p < 0.0001 | 0.924 [0.894, 0.955], p < 0.0001 | 0.922 [0.879, 0.965], p < 0.0001 | 0.936 [0.859, 1.012], p < 0.0001 |
| Lateral (x) position | -0.425 [-0.509, -0.341], p < 0.0001 | -0.478 [-0.551, -0.405], p < 0.0001 | -0.486 [-0.582, -0.391], p < 0.0001 | -0.544 [-0.738, -0.349], p < 0.0001 |
| Posterior (y) position | 0.440 [0.358, 0.522], p < 0.0001 | 0.479 [0.406, 0.552], p < 0.0001 | 0.320 [0.217, 0.423], p < 0.0001 | 0.431 [0.230, 0.631], p < 0.0001 |
| Superior (z) position | 0.284 [0.196, 0.372], p < 0.0001 | 0.303 [0.225, 0.381], p < 0.0001 | 0.228 [0.123, 0.333], p < 0.0001 | 0.075 [-0.146, 0.297], p = 0.5 |
| Spin α (°) | 0.032 [-0.060, 0.124], p = 0.5 | -0.039 [-0.121, 0.044], p = 0.4 | -0.003 [-0.111, 0.106], p = 1 | 0.161 [-0.065, 0.387], p = 0.2 |
| Verticality β (°) | -0.538 [-0.616, -0.461], p < 0.0001 | -0.533 [-0.604, -0.463], p < 0.0001 | -0.450 [-0.547, -0.354], p < 0.0001 | -0.423 [-0.630, -0.215], p = 0.0001 |
| Tilt γ (°) | 0.389 [0.302, 0.475], p < 0.0001 | 0.379 [0.303, 0.455], p < 0.0001 | 0.351 [0.249, 0.452], p < 0.0001 | 0.214 [-0.010, 0.437], p = 0.06 |

The values represent regression coefficient [95% confidence interval], two-sided p-value of each parameter with age and then body mass index (BMI). Regression coefficients were normalised by the standard deviation of both the dependent and independent variables. Cells are shaded red if the p-value is below 0.05, i.e., statistically significant with the null hypothesis that the regression coefficient is zero. LV: left ventricular. Age primarily affected the size of the heart, with increasing age reducing both cavity volume and LV mass in most subpopulations. Increasing age additionally moved the heart to a more inferior (lower z) position and made the orientation less vertical (more negative β) in some subpopulations. Most geometrical parameters were more strongly associated with BMI than age. As BMI increased, the heart centre moved to a more medial (lower x), posterior (higher y), and superior (higher z) position. Cardiac orientation also became more horizontal (more negative β) with increasing BMI. BMI did not significantly affect the cavity volume, but did make the LV mass larger, suggesting that the heart wall became thicker.